\documentclass[singlecolumn,10pt,aps,prd,superscriptaddress,preprintnumbers,amsmath,amssymb,nofootinbib]{revtex4-2}

\usepackage{epsfig}  
\usepackage{graphicx}
\usepackage{color}
\usepackage[dvipsnames]{xcolor}
\usepackage{float}
\usepackage{amsfonts}
\usepackage{amsmath}
\usepackage{mathrsfs}
\usepackage{slashed}
\usepackage{soul}
\usepackage{braket}
\usepackage{dsfont}
\usepackage[normalem]{ulem}
\usepackage{hyperref}

\begin{document}

\title{\boldmath Coherent dynamics of flavor mode entangled neutrinos}

\author{\mbox{Ashutosh Kumar Alok}}
\thanks{Deceased}
\affiliation{Indian Institute of Technology Jodhpur, Jodhpur 342037, India}

\author{\mbox{Massimo Blasone}}
\email{blasone@sa.infn.it}
\affiliation{Dipartimento di Fisica, Universitá di Salerno, Via Giovanni Paolo II, 132 I-84084 Fisciano (SA), Italy}
\affiliation{INFN, Sezione di Napoli, Gruppo collegato di Salerno, Italy}

\author{\mbox{Trambak Jyoti Chall}}
\email{chall.1@iitj.ac.in}
\affiliation{Indian Institute of Technology Jodhpur, Jodhpur 342037, India}


\author{\mbox{Neetu Raj Singh Chundawat}}
\email{chundawat.1@iitj.ac.in}
\affiliation{Indian Institute of Technology Jodhpur, Jodhpur 342037, India}
\affiliation{Institute of High Energy Physics, Chinese Academy of Sciences, Beijing 100049, China}
\affiliation{Kaiping Neutrino Research Center, Guangdong 529386, China}

\author{\mbox{Gaetano Lambiase}}
\email{lambiase@sa.infn.it}
\affiliation{Dipartimento di Fisica, Universitá di Salerno, Via Giovanni Paolo II, 132 I-84084 Fisciano (SA), Italy}
\affiliation{INFN, Sezione di Napoli, Gruppo collegato di Salerno, Italy}

\begin{abstract}
As the lynchpin of all quantum correlations, quantum coherence is fundamental for distinguishing quantum systems from classical ones and is essential for realizing quantum advantages in areas such as computation, communication, and metrology. In this study, we investigate the relationship between quantum coherence and neutrino oscillations within the two and three flavor-mode qubit frameworks. Our analysis extends beyond the commonly used \(l_1\)-norm and relative entropy of coherence to include all relevant measures of coherence such as robustness of coherence, coherence concurrence, trace-norm distance measure of coherence, coherence of formation, Schatten-\(p\)-norm-based functionals, geometric coherence and logarithmic coherence rank, each offering unique insights into the quantum correlations in these systems. Notably, while the \(l_1\)-norm and relative entropy-based measures apply to general quantum states, the other measures are particularly relevant for entangled systems, highlighting the critical role of entanglement in neutrino oscillations. We present a detailed methodology for calculating coherence measures in both two-flavor and three-flavor mixing scenarios, contributing to a deeper understanding of how quantum coherence manifests and evolves in mode-entangled neutrino systems. Our findings emphasize the potential of these systems as robust candidates for quantum information tasks, facilitated by the weak interaction nature of neutrinos.

\end{abstract}
\maketitle

\section{Introduction}
\label{intro}

Neutrino oscillations are possible when two fundamental criteria are satisfied: first, the neutrino flavor state should be expressed as a linear combination of distinct mass eigenstates, and second, the temporal evolution of a flavor state must involve a coherent superposition of the time evolutions of the associated mass eigenstates~\cite{Bilenky:1978nj}. This coherent time evolution implies the presence of mode entanglement between the mass eigenstates constituting a flavor state \cite{Blasone:2007vw,Blasone:2007wp}. Thus, neutrino oscillations with three (or two) flavors can be studied by mapping the neutrino state as a three (or two)-mode system to a three (or two)-qubit system \cite{Alok:2014gya,Banerjee:2015mha,Dixit:2018kev,Yadav:2022grk,Molewski:2021ogs,Banerjee:2025vyh,Banerjee:2024lih,Banerjee:2025gau,Bouri:2024kcl,ElBouzaidi:2025mwh}.  Another theoretically challenging scenario is when we map the neutrino states to more generic higher dimensional quantum systems like the three-level qutrit, in the case of three-flavor mixing \cite{Balantekin:2024pwc,Jha:2022yik}. Consequently, neutrino oscillations are linked to the multi-mode entanglement of single-particle states, which can be described using flavor transition probabilities \cite{Alok:2014gya}.

Over the past few decades, this concept of  mode entanglement has been a focal point of intensive theoretical and experimental  investigations. As inter-particle interactions are generally required for conventional entanglement, mode entanglement can instead arise from the coherent superposition of different occupation modes. By providing various arguments, it was demonstrated in \cite{Gerry,Enk} that the quantum state
$
|\psi\rangle_{A,B} = |0\rangle_A\otimes |1\rangle_B + |1\rangle_A \otimes|0\rangle_B,
$
where \(|0\rangle_A\) and \(|1\rangle_A\) (and similarly \(|0\rangle_B\) and \(|1\rangle_B\)) represent states with zero and one particles in modes \(A\) and \(B\), respectively is entangled and can display 
“single-particle
nonlocality”.  In refs. \cite{mode1,mode2,mode3,mode33,mode4,mode5,mode6,mode7}, it were proposed that this state can be used for applications such as teleportation, quantum cryptography, and testing Bell inequalities. Additionally, it is also possible to perform tomography on this state. In \cite{mode7}, it was demonstrated that teleportation using a single particle can be as reliable as using two particles, thereby dispelling doubts about the validity of single-photon entanglement as a resource in quantum information.

Numerous theoretical models and experimental realizations, particularly in atom-photon systems, have validated this concept, see for e.g., \cite{mode33,mode-exp2,Lombardi2002,Babichev,mode-exp3,mode-exp4,Hessmo,mode-exp5,palffy2009,mode-exp6}. The experimental realization of teleporting a one-particle entangled qubit was achieved in \cite{Lombardi2002}. In this experiment, the qubit was represented by a mode of the electromagnetic field, where the orthogonal basis states \(|0\rangle\) and \(|1\rangle\) correspond to the vacuum state and the one-photon state, respectively. In Ref. \cite{Babichev}, a single photon entangled with the vacuum was generated by directing a photon at a symmetric beam splitter. This entangled quantum state was then used to teleport single-mode quantum states of light utilizing the quantum scissors effect \cite{QS}.
The experimental demonstration of single-photon nonlocality was carried out in \cite{Hessmo} by utilizing a double homodyne measurement to assess correlations within the Fock space defined by zero and one-photon states. In \cite{palffy2009}, a coherent control scheme was presented to generate keV single-photon entanglement in nuclear forward scattering, where the spatial separation of the entangled field modes can be achieved using  X-ray polarizers and piezoelectric fast steering mirrors.

In the context of neutrino oscillations, the principle of mode entanglement provides a framework for relating flavor oscillations to the bipartite or tripartite entanglement of single particle states. If such mode entanglement is demonstrated experimentally by spatially separating the modes \cite{Blasone:2007vw}, it could open new avenues for performing various quantum information tasks, thereby providing an alternative path to advancing quantum technologies. Neutrinos, interacting primarily through weak interactions, enable unique applications over vast distances and dense media due to negligible dissipation or scattering effects. For instance, it has been demonstrated that the teleportation fidelity of mode-entangled oscillating neutrino systems consistently exceeds the classical threshold, highlighting their potential for quantum teleportation \cite{Alok:2014gya}. Utilizing the weak interaction nature of neutrinos, such teleportation could be achieved across extensive barriers, such as mountains or even the entire Earth.

In this context, it is crucial to characterize neutrino flavor oscillations by examining various facets of quantum correlations in three (or two)-qubit as well as qutrit systems. Numerous studies have already examined quantum correlations in neutrino oscillations within the framework of two- and three-qubit along with qutrit systems, see for e.g., \cite{Blasone:2007vw,Blasone:2007wp,Alok:2014gya,Banerjee:2015mha,Dixit:2018kev,Yadav:2022grk,Molewski:2021ogs,Balantekin:2024pwc,Jha:2022yik,Blasone:2014cub,Blasone:2014jea,Ming:2020nyc,Alok:2024jsd,KumarJha:2020pke,Blasone:2021cau,Li:2021fft,Li:2022mus,Li:2022zic,Blasone:2022ete,Bittencourt:2022tcl,Bittencourt:2023asd,Balantekin:2023qvm,Konwar:2024nrd,Abdolalizadeh:2024sxp}. These studies have investigated various aspects of quantum correlations, including non-locality quantified in terms of Bell, Mermin  and Svetlichny inequalities along with the non-local advantage of quantum coherence \cite{Banerjee:2015mha,Yadav:2022grk}, entanglement quantified in terms of  von Neumann and Shannon entropies and also nonlinear witness of genuine multipartite
entanglement along with other measures such as concurrence, entanglement of formation and negativity \cite{Blasone:2007vw,Alok:2014gya,Banerjee:2015mha}, discord in terms of geometric discord \cite{Alok:2014gya}, dissension \cite{Banerjee:2015mha} and using Complete Complementarity Relations in Ref. \cite{Bittencourt:2023asd}.

Quantum coherence is the defining feature that sets quantum states apart from classical states. It represents the core attribute of quantum mechanics, rooted in the universal principle of superposition, which is inherently nonclassical. In the fields of quantum information and quantum multipartite systems, quantum coherence is crucial for quantifying the inherent quantumness in the system. It can be rigorously defined within the framework of quantum resource theory and plays a vital role in numerous quantum information and estimation protocols. Additionally, it is primarily accountable for the superiority of quantum tasks over their classical counterparts. Beyond being a crucial element in quantum information processing, quantum coherence is also pivotal in emerging fields such as quantum metrology \cite{Giovannetti:2004cas}, nanoscale thermodynamics \cite{Lostaglio:2015ohe,Narasimhachar:2015cwg}, and quantum biology \cite{Lambert:2012mrj}.

The coherence effect of a quantum state is typically attributed to the off-diagonal elements of its density matrix relative to a chosen reference basis, which is defined based on the specific physical problem at hand. While the theory of quantum coherence has long been established in the field of quantum optics, its quantification has gained significant attention only in recent years, driven by advancements in quantum information science. The first thorough framework for quantifying the coherence of quantum states within resource theories was established in the groundbreaking work of Ref. \cite{Baumgratz:2014yfv}.  This work set forth stringent criteria for a quantity to qualify as a coherence measure. They also proposed the 
$l_1$-norm and the relative entropy of coherence, as two coherence measures which are based on rigorous metrics. This foundational study has since led to the development of numerous coherence measures tailored to different physical contexts, see for e.g., \cite{Yuan:2015vso,Winter:2016bkw,Shao:2014,Rana:2016vgn,Bromley:2014gna,Streltsov:2016iow,Cui:2020,Yan:2017sgd,Xi:2019,Muthuganesan:2021wnf,Zhengjun}.

Furthermore, quantum coherence is intricately linked to various measures of quantum correlations, such as entanglement, quantum discord and non-locality, see for e.g., \cite{Ma:2016gay,Xi:2015hnb,Asboth,Streltsov,Vogel,Adesso:2016ygq,Killoran}. In particular, coherence and entanglement are inherently linked. It has been shown that nonzero coherence is both necessary and sufficient for a state to generate entanglement \cite{Streltsov}. Additionally, under certain specific conditions, coherence and entanglement can be transformed into each other \cite{Asboth,Streltsov,Vogel,Adesso:2016ygq,Killoran}. Thus, quantum coherence can be viewed as the foundational element that underpins almost all quantum correlations within the system. This fundamental relationship highlights the importance of coherence in the broader context of quantum mechanics and quantum information science.

Several studies have explored quantum coherence in neutrino oscillations from different perspectives. In \cite{Song:2018bma}, quantum coherence, quantified using the $l_1$-norm, was examined in the context of three-flavor neutrino oscillations across various reactor and accelerator neutrino experiment setups. For these setups, the scale of quantum coherence—defined as the length over which coherence measures can maintain higher values—was identified. Ref. \cite{Alok:2024xeg}  explored quantum coherence in the context of neutrino spin-flavor oscillations (SFO) involving three-flavor mixing, considering both interstellar and intergalactic magnetic fields. The analysis employed the 
$l_1$-norm and relative entropy of coherence as quantifiers. Remarkably, the findings indicated that in the SFO scenario, quantum coherence could persist over incredibly vast astrophysical distances, spanning from kiloparsecs to gigaparsecs, suggesting that coherence is robust even across cosmic scales. 

In this work, we present a comprehensive study of quantum coherence in flavor-mode entangled neutrinos by analyzing several relevant measures of quantum coherence in the context of three-flavor neutrino oscillations, where the neutrino state is mapped from a three-mode system to the tripartite basis of a three-qubit system. For the sake of completeness, we also provide a detailed methodology for the coherence measures when the neutrino states (as two-mode systems) are mapped to the bipartite basis of the two-qubit system in the case of two-flavor mixing.
Apart from the $l_1$-norm of coherence, we consider the relative entropy of coherence,   robustness of coherence, coherence concurrence, trace-norm
distance measure of coherence, coherence of formation, Schatten-$p$-norm-based functionals, geometric coherence and the logarithmic coherence rank. It is important to note that, unlike the $l_1$-norm of coherence and the relative entropy of coherence, the other three measures are valid only for entangled systems.

The structure of this paper is organized as follows. In the next section, we present the methodology for modeling neutrinos as a qubit systems, laying the groundwork for the analysis that follows. Sec. III presents the theoretical calculations of various quantum coherence measures for mode-entangled neutrinos, treating them as two- and three-qubit systems. We map these theoretical expressions to plots that illustrate the dynamic behavior of these coherence measures in Sec. IV. Finally, we provide our conclusions in Sec. V.

\section{Neutrinos as qubits }
The three neutrino flavor states, \(\nu_e\), \(\nu_{\mu}\), and \(\nu_{\tau}\), undergo oscillations due to the mismatch between the flavor and mass bases. The flavor eigenstates are related to the mass eigenstates \(\nu_1\), \(\nu_2\), and \(\nu_3\) through the unitary \(3\times3\) Pontecorvo-Maki-Nakagawa-Sakata (PMNS) leptonic mixing matrix \(U_{\alpha i}\), such that
\begin{equation}
    |\nu_{\alpha}(t)\rangle
=
\sum_{i=1}^3 U_{\alpha i}|\nu_i(t)\rangle
=
\sum_{i=1}^3 e^{-iE_it}U_{\alpha i}|\nu_i(0)\rangle.
    \label{3mix}
\end{equation}
where \(\alpha=e,\mu,\tau\), and the mass eigenstates evolve with energies \(E_i\). The PMNS matrix is given by
\begin{equation}
    U_{\alpha i}=\begin{pmatrix}
        c_{12}c_{13} & s_{12}c_{13} & s_{13} \\
        -s_{12}c_{23}-c_{12}s_{23}s_{13}  &
        c_{12}c_{23}-s_{12}s_{23}s_{13} &
        s_{23}c_{13} \\
        s_{12}s_{23}-c_{12}c_{23}s_{13} &
        -c_{12}s_{23}-s_{12}c_{23}s_{13} &
        c_{23}c_{13}
    \end{pmatrix},
    \label{pmns}
\end{equation}
where \(s_{ij}\equiv\sin\theta_{ij}\) and \(c_{ij}\equiv\cos\theta_{ij}\). The mixing is characterized by the three mixing angles \(\theta_{12}\), \(\theta_{13}\), and \(\theta_{23}\), along with the complex Charge-Parity (CP) violating phase \(\delta_{\rm CP}\). In the present work, we restrict our analysis to the CP-conserving case by taking \(\delta_{\rm CP}=0\), thereby neglecting CP-violating effects. For nonzero \(\delta_{\rm CP}\), additional complex phases enter the oscillation amplitudes and can modify the quantitative behavior of the coherence measures. However, to keep the present analysis focused on the comparison of different quantum coherence quantifiers, we restrict our analysis to the simplified case of \(\delta_{\rm CP}=0\).

In experimentally relevant scenarios, the complex three-flavor neutrino mixing can be simplified to an effective two-flavor mixing model using the SU(2) unitary rotation matrix. This reduction is expressed as
\begin{equation}
    \begin{pmatrix}
        |\nu_{\alpha}(t)\rangle \\ |\nu_{\beta}(t)\rangle
    \end{pmatrix}= \begin{pmatrix}
        \cos{\theta} & \sin{\theta} \\ -\sin{\theta} & \cos{\theta}
    \end{pmatrix} \begin{pmatrix}
         e^{-i E_i t}|\nu_{i}\rangle \\ e^{-i E_j t}|\nu_{j}\rangle 
    \end{pmatrix}\,,
    \label{mix}
\end{equation}
where $\alpha$ and $\beta$ are the flavor indices which can be $e$, $\mu$ or $\tau$, and $i$ and $j$ are the mass indices. Since in the case of two flavor mixing only one mixing angle between the two flavors is required, we denote the parameter simply by $\theta$.
In the relativistic limit, the exact quantum field theoretic flavor states reduce to the more familiar Pontecorvo flavor states. These states are physically well-defined, representing distinct single-particle entities associated with specific flavor modes. Consequently, this framework allows us to define and analyze entanglement between these flavor modes, highlighting the quantum nature of neutrino oscillations.

The two-level ``qubit'' system is described by the two mutually orthogonal states 
\[
|0\rangle = \begin{pmatrix} 
1 \\ 
0 
\end{pmatrix} \quad \text{and} \quad 
|1\rangle = \begin{pmatrix} 
0 \\ 
1 
\end{pmatrix},
\] 
which exist within a complex two-dimensional Hilbert space \(\mathcal{H}^2\). In this framework, we adopt the occupation number representation, where the presence of a single-particle state in a specific flavor mode (say \(\alpha\)) is denoted by the qubit \( |1\rangle_{\alpha} \), and its absence by \( |0\rangle_{\alpha} \).
Using this scheme, the flavor states at \( L = 0 \) (in ultrarelativistic limits, $t \sim L$) can be mapped to the bipartite basis states in a two-qubit system (for two-flavor mixing) as 
\[
|\nu_{\alpha}\rangle = |1\rangle_{\alpha} \otimes |0\rangle_{\beta} \quad \text{and} \quad 
|\nu_{\beta}\rangle = |0\rangle_{\alpha} \otimes |1\rangle_{\beta}.
\]
Similarly, for three-flavor mixing, the flavor states at \( L = 0 \) can be mapped to tripartite states in a three-qubit system as follows:
\begin{eqnarray}
|\nu_e\rangle = |1\rangle_e \otimes |0\rangle_{\mu} \otimes |0\rangle_{\tau},\nonumber\\
|\nu_{\mu}\rangle = |0\rangle_e \otimes |1\rangle_{\mu} \otimes |0\rangle_{\tau},\nonumber \\
|\nu_{\tau}\rangle = |0\rangle_e \otimes |0\rangle_{\mu} \otimes |1\rangle_{\tau}.
\end{eqnarray}
This approach effectively captures the quantum states of neutrinos in terms of qubits, allowing for a clear representation of flavor mixing within two- and three-flavor systems.

Therefore, in the case of two-flavor mixing, the evolution of a single-particle mode-entangled neutrino flavor state, which starts in the flavor \(\alpha\) (where \(\alpha\) can be \(e\), \(\mu\), or \(\tau\)), can be described as a Bell-like superposition of two qubits in the Hilbert space \(\mathcal{H}^2 \otimes \mathcal{H}^2\) as
\begin{equation}
|\nu_{\alpha}(\theta,L)\rangle = \Tilde{U}_{\alpha\alpha}(\theta,L) |1\rangle_{\alpha} \otimes |0\rangle_{\beta} + \Tilde{U}_{\alpha\beta}(\theta,L) |0\rangle_{\alpha} \otimes |1\rangle_{\beta},
\label{2evol}
\end{equation}
where \(\alpha\) and \(\beta\) represent any two of the three available flavor modes: \(e\), \(\mu\), or \(\tau\). The coefficients of this Bell-like superposition, derived from the parameters governing neutrino flavor oscillations, are expressed as:
\begin{eqnarray}
\Tilde{U}_{\alpha\alpha}(\theta,L) &=& \cos^2{\theta} e^{-iE_i L} + \sin^2{\theta} e^{-iE_j L}, \nonumber\\
\Tilde{U}_{\alpha\beta}(\theta,L) &=& \cos{\theta} \sin{\theta} \left(e^{-iE_j L} - e^{-iE_i L}\right),
\label{2coef}
\end{eqnarray}
where \(i, j = 1, 2, 3\) are the mass indices. The normalization condition is \( |\Tilde{U}_{\alpha\alpha}(\theta,L)|^2 + |\Tilde{U}_{\alpha\beta}(\theta,L)|^2 = 1 \), reflecting that \( |\Tilde{U}_{\alpha\alpha}(\theta,L)|^2 \) represents the survival probability of the flavor \(\alpha\), while \( |\Tilde{U}_{\alpha\beta}(\theta,L)|^2 \) represents the probability of transition from the initial flavor \(\alpha\) to \(\beta\).
\begin{widetext}

Now from Eq. \eqref{2evol}, the $4\times 4$ two-qubit density matrix $\rho_{\alpha,2}^{4\times 4}(\theta,L)=\ket{\nu_{\alpha}(\theta,L)}\bra{\nu_{\alpha}(\theta,L)}$ for two neutrino flavor mixing can be written  as\footnote{The Dirac matrices which are tensor products of the Pauli matrices (that generate the $\rm SU(2)$ group) $D_{\mu\nu}=\sigma_{\mu}\otimes\sigma_{\nu}$ serve as the basis for constructing these $4\times 4$ density matrices.},
\begin{equation} 
    \rho_{\alpha,2}^{4\times 4}(\theta,L)=\begin{pmatrix}
        0&0&0&0\\0& |\Tilde{U}_{\alpha\alpha}(\theta,L)|^2 & \Tilde{U}_{\alpha\alpha}(\theta,L)\Tilde{U}_{\alpha\beta}^{\ast}(\theta,L) &0\\0& \Tilde{U}_{\alpha\alpha}^{\ast}(\theta,L)\Tilde{U}_{\alpha\beta}(\theta,L) & |\Tilde{U}_{\alpha\beta}(\theta,L)|^2 &0\\0&0&0&0
    \end{pmatrix}\,.
    \label{den2}
\end{equation}
The time evolution of a mode-entangled neutrino flavor state in the three-flavor mixing scenario can be represented as a superposition of three qubits within the Hilbert space \(\mathcal{H}^2 \otimes \mathcal{H}^2 \otimes \mathcal{H}^2\) as
\footnote{We get this symmetric transformation matrix due to the fact that we consider negligible $CP$-violating phase in the mixing matrix, and hence all the PMNS matrix elements are considered to be real, which results in the coefficients in Eq. \eqref{a} being symmetric in the two indices $\alpha,\beta$.}\cite{Banerjee:2015mha},

\begin{equation}
    \begin{pmatrix}
        |\nu_{e}(\theta_{ij},L)\rangle\\|\nu_{\mu}(\theta_{ij},L)\rangle\\|\nu_{\tau}(\theta_{ij},L)\rangle
    \end{pmatrix}=\begin{pmatrix}
        a_{ee}\left(\theta_{ij},L\right) & a_{e\mu}\left(\theta_{ij},L\right) & a_{e\tau}\left(\theta_{ij},L\right) \\a_{e\mu}\left(\theta_{ij},L\right) & a_{\mu\mu}\left(\theta_{ij},L\right) & a_{\mu\tau}\left(\theta_{ij},L\right) \\ a_{e\tau}\left(\theta_{ij},L\right) & a_{\mu\tau}\left(\theta_{ij},L\right) & a_{\tau\tau}\left(\theta_{ij},L\right)
    \end{pmatrix} \begin{pmatrix}
        |1\rangle\otimes |0\rangle\otimes |0\rangle\\|0\rangle\otimes |1\rangle\otimes |0\rangle\\|0\rangle\otimes |0\rangle\otimes |1\rangle
    \end{pmatrix}\,,
    \label{3evol2}
\end{equation}
where the complex coefficients $a_{\alpha\beta}\left(\theta_{ij},L\right)$ with $\left(\alpha,\beta\right)=e,\mu,\tau$ are given as,
\begin{eqnarray}
a_{ee}\left(\theta_{ij},L\right)&=&c_{12}^2c_{13}^2 e^{-i E_1 L} + s_{12}^2c_{13}^2 e^{-i E_2 L}+ s_{13}^2 e^{-i E_3 L},\nonumber\\
a_{\mu\mu}\left(\theta_{ij},L\right)&=&(s_{12}c_{23}+c_{12}s_{23}s_{13})^2 e^{-i E_1 L} + (c_{12}c_{23}-s_{12}s_{23}s_{13})^2 e^{-i E_2 L}+ s_{23}^2c_{13}^2 e^{-i E_3 L},\nonumber\\
a_{\tau\tau}\left(\theta_{ij},L\right)&=&(s_{12}s_{23}-c_{12}c_{23}s_{13})^2 e^{-i E_1 L} + (c_{12}s_{23}+s_{12}c_{23}s_{13})^2 e^{-i E_2 L}+ c_{23}^2c_{13}^2 e^{-i E_3 L},\nonumber\\
a_{e\mu}\left(\theta_{ij},L\right)&=& -(c_{12}c_{13}s_{12}c_{23}+c_{12}^2c_{13}s_{23}s_{13}) e^{-i E_1 L} + (s_{12}c_{13}c_{12}c_{23}-s_{12}^2c_{13}s_{23}s_{13})e^{-i E_2 L}\nonumber\\&+&\quad s_{13}s_{23}c_{13} e^{-i E_3 L},\nonumber\\
a_{e\tau}\left(\theta_{ij},L\right)&=& (c_{12}c_{13}s_{12}s_{23}-c_{12}^2c_{13}c_{23}s_{13}) e^{-i E_1 L} -(s_{12}c_{13}c_{12}s_{23}+s_{12}^2c_{13}c_{23}s_{13})e^{-i E_2 L}\nonumber\\&+&\quad  s_{13}c_{23}c_{13} e^{-i E_3 L},\nonumber\\
a_{\mu\tau}\left(\theta_{ij},L\right)&=&
    (-s^2_{12}c_{23}s_{23} +s_{12}c_{23}^2c_{12}s_{13} -c_{12}s_{23}^2s_{13}s_{12}+c_{12}^2s_{23}s_{13}^2c_{23})e^{-i E_1 L}\nonumber\\
    &+&(-c_{12}^2c_{23}s_{23}-c_{12}c_{23}^2s_{12}s_{13}+s_{12}s_{23}^2s_{13}c_{12}+s_{12}^2c_{23}s_{13}^2s_{23})e^{-i E_2 L}\nonumber\\
    &+&\quad s_{23}c_{13}^2c_{23} e^{-i E_3 L}\,.
    \label{a}
\end{eqnarray}
The decompositions in Eq. \eqref{3evol2} in the tripartite three-qubit basis gives us three $8\times 8$ three-qubit density matrices $\rho_{\alpha,3}^{8\times 8}(\theta_{ij},L)=\ket{\nu_{\alpha}(\theta_{ij},L)} \bra{\nu_{\alpha}(\theta_{ij},L)}$ for the three initial flavors $\alpha=e,\mu,\tau$ as,
\begin{equation}
    \rho_{\alpha,3}^{8\times 8}(\theta_{ij},L)=\begin{pmatrix}
0&0&0&0&0&0&0&0\\0&|a_{\alpha\tau}\left(\theta_{ij},L\right)|^2&a_{\alpha\tau}^{\ast}\left(\theta_{ij},L\right)a_{\alpha\mu}\left(\theta_{ij},L\right)&0&a_{\alpha\tau}^{\ast}\left(\theta_{ij},L\right)a_{\alpha e}\left(\theta_{ij},L\right)&0&0&0\\0&a_{\alpha\mu}^{\ast}\left(\theta_{ij},L\right)a_{\alpha\tau}\left(\theta_{ij},L\right)&|a_{\alpha\mu}\left(\theta_{ij},L\right)|^2&0&a_{\alpha\mu}^{\ast}\left(\theta_{ij},L\right)a_{\alpha e}\left(\theta_{ij},L\right)&0&0&0\\0&0&0&0&0&0&0&0 \\0&a_{\alpha e}^{\ast}\left(\theta_{ij},L\right)a_{\alpha\tau}\left(\theta_{ij},L\right)&a_{\alpha e}^{\ast}\left(\theta_{ij},L\right)a_{\alpha\mu}\left(\theta_{ij},L\right)&0&|a_{\alpha e}\left(\theta_{ij},L\right)|^2&0&0&0\\0&0&0&0&0&0&0&0\\0&0&0&0&0&0&0&0\\0&0&0&0&0&0&0&0
    \end{pmatrix}\,.
    \label{den32}
\end{equation}
\end{widetext}

One may also represent the three-flavour neutrino system spanned by three mutually orthogonal states residing in a complex three-dimensional Hilbert space \( \mathcal{H}^3 \). However, when this qutrit basis is chosen in the flavor basis the resulting coherence measures become equivalent to the three-qubit flavor-mode representation due to the isometric correspondence
\begin{equation}
|\nu_e\rangle \leftrightarrow |100\rangle,\qquad
|\nu_\mu\rangle \leftrightarrow |010\rangle,\qquad
|\nu_\tau\rangle \leftrightarrow |001\rangle.
\end{equation}

Hence, the flavor-basis qutrit and the single-excitation three-qubit representations describe the same physical state. Consequently, the coherence measures such as the \(l_1\)-norm of coherence, relative entropy of coherence, coherence concurrence and robustness of coherence becomes identical for these two representations. Therefore, in the subsequent sections, we focus primarily on the three-qubit flavor-mode framework. We are now equipped with all the density matrices required for the calculation of various coherence measures in both the two-flavor and three-flavor mixing scenarios.

\section{Measures of quantum coherence}
A completely positive and trace preserving ($\rm CPTP$) $\rm U(1)$-covariant incoherent operation can be expressed as a function of the density operator $\rho$ as, 
\begin{equation} 
    \Xi_{\rm ICPTP}(\rho) = \sum_{\ell} K_\ell \rho K_\ell^\dagger\,, 
\end{equation} 
where $K_\ell$ represents the incoherent Kraus operators that satisfy $\sum_{\ell}K_\ell^\dagger K_\ell=\mathds{1}$ and $\ell$ represents individual outcomes corresponding to which the states are given by $\rho_{\ell}=\frac{K_{\ell}\rho K_{\ell}^{\dagger}}{p_{\ell}}$ where the probability of occurrence of such a state is given by $p_{\ell}={\rm Tr}(K_\ell \rho K_\ell^\dagger)$.
The Kraus operators transform any incoherent state into another incoherent state fulfilling $K_\ell \mathcal{I}  K_\ell^\dagger\, \subset \mathcal{I}$ $ \forall$ $\ell$, with $ \mathcal{I}$
representing the set of incoherent quantum states which can be expressed as
\(
\rho=\sum_{i=1}^{d} p_i \ket{i}\bra{i},
\)
where \(p_i\) are probabilities and the Hilbert space \(\mathcal{H}^d\) is \(d\)-dimensional. So the states diagonal in the reference basis $\{\ket{i}\}_{\rm i=1,2,..n}$ are incoherent, and the states that cannot be expressed in this form are termed coherent. 
Coherence can be mathematically depicted as a functional of density operators $C(\rho)$ that maps the states to non-negative real numbers. This mapping has to satisfy some conditions for the measure of coherence to be called ``proper''.

The first condition (\(\boldsymbol{\rm C_1}\)) is \textit{Faithfulness}, which asserts that \( C(\rho) \geq 0 \) for any quantum state \(\rho\), and specifically, \( C(\rho) \) should vanish for incoherent states. In other words, \( C(\sigma) = 0 \) if and only if \(\sigma \in \mathcal{I}\), meaning that \(\sigma\) is an incoherent state. This condition is crucial because it ensures that the coherence measure accurately distinguishes between coherent and incoherent states. This condition is also often referred to as the ``Non-negativity'' condition, reflecting its requirement that the measure be non-negative for all states.

The second condition is \textit{Monotonicity} (\(\boldsymbol{\rm C_2}\)), which requires that the coherence measure \( C(\rho) \) should not increase under incoherent completely positive and trace-preserving (CPTP) maps. Mathematically, this is expressed as \( C(\Xi_{\rm ICPTP}(\rho)) \leq C(\rho) \) for all incoherent operations \(\Xi_{\rm ICPTP}\). This condition ensures that coherence, as a resource, cannot be generated for free through operations that are deemed incoherent. However, this condition does not account for scenarios where the measurement outcomes are retained. To capture these scenarios, a stronger condition, known as \textit{Strong Monotonicity} (\(\boldsymbol{\rm C_2'}\)), is imposed. This condition requires that \( C(\rho) \geq \sum_{\ell} p_{\ell} C(\rho_{\ell}) \) for all sets of Kraus operators \(\{K_{\ell}\}\), ensuring that the expected coherence does not increase even when the measurement results are known. This stronger requirement is vital, as it rules out certain measures, such as those based on fidelity, from being considered proper measures of coherence since they do not generally satisfy \(\rm C_2'\)~\cite{Shao:2014}.

The final essential condition on \( C(\rho) \) is \textit{Convexity} (\(\boldsymbol{\rm C_3}\)), which stipulates that the coherence measure should not increase under the mixing of quantum states. Specifically, for any set of states \(\{\rho_{\ell}\}\) and probabilities \(p_{\ell} \geq 0\) with \(\sum_{\ell} p_{\ell} = 1\), the condition requires that \( C(\sum_{\ell} p_{\ell} \rho_{\ell}) \leq \sum_{\ell} p_{\ell} C(\rho_{\ell}) \). This condition is important because it prevents artificial increases in coherence due to probabilistic mixtures of states, thereby ensuring that the measure respects the notion that mixing cannot generate coherence.

We will now explore  ten potential coherence monotones within the context of two- and three-flavor mode-entangled neutrino oscillations. In these cases, the neutrino states are mapped onto two-qubit systems for two-flavor mixing and three-qubit system for three-flavor mixing. We will express these monotones in terms of measurable parameters associated with neutrino oscillations.

\subsection{$l_1$-norm of coherence}

If $\mathcal{V}^{m\times n}$ is a vector space with $m\times n$ matrices, then the $l_{p}$ norm on $\mathcal{V}^{m\times n}$ induced by the vector $p$-norms for both the vector spaces $\mathcal{V}^{m}$ and $\mathcal{V}^{n}$ can be given as $\left(\sum_{m\neq n}|\rho_{mn}|^{p}\right)^{\frac{1}{p}}$, where $\rho_{mn}=\langle m|\rho|n\rangle$ are the matrix elements~\cite{Rana:2016vgn}. For $p=1$, the $l_{p=1}$ matrix norm satisfies all the conditions $\boldsymbol{\rm C_1},\boldsymbol{\rm C_2},\boldsymbol{\rm C_2'}$ and $\boldsymbol{\rm C_3}$ for being a proper measure of coherence, and clasps the sum of the absolute magnitudes of all off-diagonal entries in the density matrix $\rho(t)$ and can take a maximum value of $r-1$, where $r$ represents the dimension of $\rho(t)$.
Hence, as a function of $\rho(t)$, the \textit{$l_{1}$-norm of coherence} is given as,
\begin{equation}
C_{l_1}\left(\rho\right)\equiv\sum_{m\neq n}\big|\langle m|\rho(t)|n\rangle\big|\geq 0\,.
     \label{coh}
\end{equation}

The flavor-mode qubit representation possesses a different off-diagonal density matrix structure, which can lead to different quantitative values of the coherence measures as compared to the standard description of neutrino oscillations. Now let us first consider the effective two-flavor mixing scenario, where we map the initial neutrino flavor states to the bipartite basis states in the two-qubit coupled system represented by the $4\times 4$ density matrix $\rho_{\alpha,2}^{4\times 4}(\theta,L)$ in Eq. \eqref{den2} using which, the $l_{1}$-norm of coherence can be expressed through the definition in Eq. \eqref{coh} as
\begin{eqnarray}
C_{l_1}^{2f}\left(\rho_{\alpha,2}^{4\times 4}\right)&=&\sum_{m\neq n}\big|\langle m|\rho_{\alpha,2}^{4\times 4}(\theta,L)|n\rangle\big|\nonumber\\&=&\big|\Tilde{U}_{\alpha\alpha}(\theta,L)\Tilde{U}_{\alpha\beta}^{\ast}(\theta,L)\big| + \big|\Tilde{U}_{\alpha\alpha}^{\ast}(\theta,L)\Tilde{U}_{\alpha\beta}(\theta,L)\big|\,.\nonumber\\
     \label{2l1-1}
\end{eqnarray}
Now we can use the coefficients from Eq. \eqref{2coef} in the above equation, which gives us,
\begin{equation}
\begin{aligned}
C_{l_1}^{2f}\left(\rho_{\alpha,2}^{4\times 4}\right) &= \Big\{
\big| (\cos^2\theta e^{-iE_i L} + \sin^2\theta e^{-iE_j L})\sin\theta\cos\theta (e^{iE_j L}-e^{iE_i L}) \big| \\
&\quad + \big|(\cos^2\theta e^{iE_i L} + \sin^2\theta e^{iE_j L}) \sin\theta\cos\theta (e^{-iE_j L}-e^{-iE_i L}) \big|\Big\} \\
&= 2|\sin\theta\cos\theta| \big|e^{iE_j L}-e^{iE_i L}\big| \big|\cos^2\theta e^{iE_i L} + \sin^2\theta e^{iE_j L}\big|\,.
\end{aligned}
\end{equation}

On simplification, we arrive at
\begin{eqnarray}     
     C_{l_1}^{2f}\left(\rho_{\alpha,2}^{4\times 4}\right)&=&  |\sin{2\theta}|\sqrt{2\left(1-\cos{\frac{\Delta m_{ij}^{2}}{2E}L}\right)\sin^2{\theta}\cos^2{\theta}\left(\cot^2{\theta}+\tan^2{\theta}+2\cos{\frac{\Delta m_{ij}^{2}}{2E}L}\right)} \nonumber\\
     &=& \frac{\sin^2{2\theta}}{\sqrt{2}}\sqrt{\cot^2{\theta}+\tan^2{\theta}-2\cos^2{\frac{\Delta m_{ij}^{2}}{2E}L}-\big(\cot{\theta}-\tan{\theta}\big)^2\cos{\frac{\Delta m_{ij}^{2}}{2E}L}} \nonumber \,
     \\&=&
    2\sqrt{P_{\alpha\alpha}^{2f}P_{\alpha\beta}^{2f}}\,,
      \label{2l1}
     \end{eqnarray}
where \(P_{\alpha\alpha}^{2f}=1-P_{\alpha\beta}^{2f}\) and
\(
P_{\alpha\beta}^{2f}
=
\sin^2 2\theta\,
\sin^2\!\left(\frac{\Delta m_{ij}^{2}L}{4E}\right),
\)
are survival and oscillation probabilities, respectively. We use the relationship \(E_i-E_j=\frac{\Delta m_{ij}^2}{2E}\) for \(i > j\), where \(E_i\) and \(E_j\) are the energy eigenvalues corresponding to the mass eigenstates \(\nu_i\) and \(\nu_j\), respectively. This relationship is crucial for expressing the phase differences in terms of the mass-squared differences \(\Delta m_{ij}^{2}\), which are key measurable parameters in neutrino oscillation experiments.
The mixing angle \(\theta\) between the two states \(i\) and \(j\) (where \(i, j = 1, 2, 3\)) is denoted simply as \(\theta\), and it can correspond to any of the mixing angles \(\theta_{12}\), \(\theta_{23}\), or \(\theta_{13}\) depending on the specific flavors involved in the mixing. This notation is consistent with the previous sections, where we used \(\theta\) to describe the mixing between different flavor states. For the two-flavor mixing scenario, we will continue using this \(\theta\)-notation for the calculation of the remaining coherence measures, ensuring consistency and clarity throughout the analysis.


When we map the neutrino states to the tripartite three-qubit basis, considering the three-flavor evolution of the mode entangled neutrino  $|\nu_{\alpha}\left(\theta_{ij},L\right)\rangle$ of an initial flavor $\alpha$, we use the $8\times 8$ density matrices $\rho_{\alpha,3}^{8\times 8}(\theta_{ij},L)$ from Eq. \eqref{den32} and get the $l_{1}$-norm of coherence as,

\begin{equation}
\begin{aligned}
C_{l_1}^{3f}\left(\rho_{\alpha,3}^{8\times 8}\right) 
&= \sum_{m\neq n}\big|\langle m|\rho_{\alpha,3}^{8\times 8}(\theta_{ij},L)|n\rangle\big|  \\
&= \frac{1}{2}\sum_{\substack{\beta,\gamma=e\\ \beta\neq\gamma}}^{\tau}
\Big(
\big|a_{\alpha\beta}^{\ast}(\theta_{ij},L)a_{\alpha\gamma}(\theta_{ij},L)\big|
+
\big|a_{\alpha\beta}(\theta_{ij},L)a_{\alpha\gamma}^{\ast}(\theta_{ij},L)\big|
\Big)\\
&= \sum_{\substack{\beta,\gamma=e\\ \beta\neq\gamma}}^{\tau}
\big|a_{\alpha\beta}^{\ast}(\theta_{ij},L)a_{\alpha\gamma}(\theta_{ij},L)\big|
\label{3l12}
\end{aligned}
\end{equation}

where the complex coefficients $a_{\alpha(e,\mu,\tau)}\left(\theta_{ij},L\right)$ from Eq. \eqref{a} can be put into Eq. \eqref{3l12} to express $C_{l_1}^{3f}\left(\rho_{\alpha,3}^{8\times 8}\right)$ in terms of the measurable neutrino oscillation parameters.

The $l_1$-norm of coherence considering the initial neutrino flavor $\alpha=e$, i.e. $C_{l_1}^{3f}\left(\rho_{e,3}^{8\times 8}\right)$ can be expressed in terms of the measurable parameters as,
\begin{equation}
\begin{aligned}
C_{l_1}^{3f}\left(\rho_{e,3}^{8\times 8}\right) &= 2\Bigl\{\left|a_{ee}^{\ast}(\theta_{ij},L)a_{e\mu}(\theta_{ij},L)\right| + \left|a_{ee}^{\ast}(\theta_{ij},L)a_{e\tau}(\theta_{ij},L)\right| + \left|a_{e\mu}^{\ast}(\theta_{ij},L)a_{e\tau}(\theta_{ij},L)\right|\Bigr\}.
\label{l1-e3}
\end{aligned}
\end{equation}

The $l_1$-norm of coherence considering the initial muon neutrino flavor $\alpha=\mu$, i.e. $C_{l_1}^{3f}\left(\rho_{\mu,3}^{8\times 8}\right)$ can be written as,

\begin{equation}
\begin{aligned}
C_{l_1}^{3f}\left(\rho_{\mu,3}^{8\times 8}\right) = 2\Bigl\{
 \left|a_{e\mu}^{\ast}(\theta_{ij},L)a_{\mu\mu}(\theta_{ij},L)\right| 
 + \left|a_{e\mu}^{\ast}(\theta_{ij},L)a_{\mu\tau}(\theta_{ij},L)\right| 
 + \left|a_{\mu\mu}^{\ast}(\theta_{ij},L)a_{\mu\tau}(\theta_{ij},L)\right|\Bigr\}.
\end{aligned}
\end{equation}

Using the relation
\(
P_{\alpha\beta}(\theta_{ij},L)
=
|a_{\alpha\beta}(\theta_{ij},L)|^2,
\)
where the three-flavor oscillation probabilities are given by
\begin{equation}
\begin{aligned}
P_{\alpha\beta}(\theta_{ij},L)
=
\delta_{\alpha\beta}
-
4\sum_{j>i=1}^{3}
\mathrm{Re}\!\left[
U_{\alpha j}^{\ast}U_{\beta j}
U_{\alpha i}U_{\beta i}^{\ast}
\right]
\sin^2\!\left(\frac{\Delta m_{ji}^{2}L}{4E}\right),
\label{moda2}
\end{aligned}
\end{equation}
with \(U_{\alpha i}\) (\(\alpha=e,\mu,\tau\), \(i=1,2,3\)) denoting the PMNS matrix elements defined in Eq.~\eqref{pmns}, the expression in Eq.~\eqref{3l12} can be recast entirely in terms of measurable oscillation probabilities as
\begin{equation}
\begin{aligned}
C_{l_1}^{3f}\left(\rho_{\alpha,3}^{8\times 8}\right)
=
2\Big[
&\sqrt{P_{\alpha e}(\theta_{ij},L)\,P_{\alpha\mu}(\theta_{ij},L)}
+\sqrt{P_{\alpha e}(\theta_{ij},L)\,P_{\alpha\tau}(\theta_{ij},L)}
+\sqrt{P_{\alpha\mu}(\theta_{ij},L)\,P_{\alpha\tau}(\theta_{ij},L)}
\Big].
\label{eq:l1prob}
\end{aligned}
\end{equation}

\noindent \textbf{Robustness of coherence}: The robustness of coherence \( C_R(\rho) \) is another important measure for quantifying coherence in quantum systems. This measure is defined within the convex set of density operators \( \mathcal{D}(\mathbb{C}^d) \), which act on a \( d \)-dimensional Hilbert space, as detailed in \cite{Napoli:2016vpc}. Mathematically, it is expressed as:
\begin{equation}
C_R\left(\rho\right)=\min_{\tau\in\mathcal{D}\left(\mathbb{C}^{d}\right)} \left\{ s\geq0\left| \frac{\rho+s\tau}{1+s}=:\delta\in\mathcal{I} \right. \right\}
\end{equation}
This equation defines the robustness of coherence as the minimal weight \( s \) of another arbitrary quantum state \( \tau \) that, when mixed convexly with the state \( \rho \), produces an incoherent state \( \delta \) belonging to the set of incoherent states \( \mathcal{I} \). In essence, this measure quantifies the smallest amount of "noise" or additional state \( \tau \) required to make the quantum state \( \rho \) indistinguishable from an incoherent state. The robustness of coherence, therefore, provides a clear operational meaning: it measures the resilience of quantum coherence against the introduction of external disturbances. This resilience is particularly relevant in practical scenarios where the integrity of quantum coherence is crucial, such as in quantum computation, quantum communication, and other quantum information processing tasks. By understanding the robustness of coherence, one can gauge how much external noise a quantum system can tolerate before losing its coherence, thereby offering insights into the stability and reliability of the quantum states under various conditions.

This measure is particularly useful because it can also be characterized by linear inequalities in the case of a \(d\)-dimensional quantum system, as shown in \cite{Piani:2016lse}. Specifically, the robustness of coherence satisfies the following bounds:
\begin{equation}
    \frac{C_{l_1}\left(\rho\right)}{d-1} \leq C_R\left(\rho\right) \leq C_{l_1}\left(\rho\right),
\end{equation}
where both the upper and lower bounds are quite tight. Notably, the upper bound becomes an equality, i.e., \( C_R\left(\rho\right) = C_{l_1}\left(\rho\right) \), for pure states. This equality is particularly relevant for the neutrino mode-entangled systems under consideration, where the robustness of coherence simplifies to the \( l_1 \)-norm of coherence, thereby offering a straightforward method for analyzing coherence in these systems.

\noindent \textbf{Trace-norm distance measure of coherence}: A measure of coherence induced by the trace-norm distance of the quantum state $\rho$ to the nearest incoherent state $\delta$ can be given as,
    \begin{equation}
        C_{tr}\left(\rho\right)=\min_{\delta\epsilon\mathcal{I}} D\left(\rho,\delta\right)=\min_{\delta\epsilon\mathcal{I}}\left\{{\rm Tr} |\rho-\delta|\right\}\,,
    \end{equation}
where $D\left(\rho,\delta\right)={\rm Tr} |\rho-\delta|$ is the trace-norm distance between $\rho$ and $\delta$.  A smaller \(C_{\text{tr}}(\rho)\) value implies that the state is closer to the nearest incoherent state, and hence, has less coherence. Conversely, a larger value indicates that the state is further away from the incoherent state, implying a higher degree of quantum coherence.
It is still unknown whether $C_{tr}$ obeys the strong monotonicity condition $\boldsymbol{\rm C_2'}$ \textit{in general}. 

However, for two-qubit Bell-like pure states, such as those encountered in the two-flavor mode-entangled neutrino case, the trace norm of coherence \( C_{tr} \) coincides with the \( l_1 \)-norm of coherence \( C_{l_1} \), thereby making it a proper measure of coherence in this context\footnote{It is important to note that this equivalence between \( C_{tr} \) and \( C_{l_1} \) does not extend to general two-qubit states, where the relationship between these measures can differ significantly.}\cite{Rana:2016vgn,Bromley:2014gna}. As a result, for the two-flavor mixing scenario involving two-qubit neutrinos, the trace norm of coherence \( C_{tr}^{2f}\left(\rho_{\alpha,2}\right) \) is effectively equal to the \( l_1 \)-norm of coherence \( C_{l_1}^{2f}\left(\rho_{\alpha,2}\right) \), which can be computed using the expression provided in Eq. \eqref{2l1}. For pure qubit states \( \ket{\psi}\bra{\psi} \), the nearest incoherent state is simply \( \delta = \rho_D = \text{diag}\left\{\ket{\psi}\bra{\psi}\right\} \), where \( \rho_D \) is the dephased state. However, this straightforward identification does not hold for states with dimensions higher than 2, i.e., for \( d > 2 \).

\noindent \textbf{Coherence concurrence}: A relatively new and rigorous measure of quantum coherence, known as coherence concurrence, was introduced in \cite{Yan:2017sgd} to explore the interplay and interconversion between two fundamental aspects of quantumness: entanglement and coherence. This measure provides a deeper understanding of the relationship between these two critical quantum resources, especially in systems where both coherence and entanglement play significant roles. The coherence concurrence is particularly relevant in the study of complex quantum systems, where it serves as a valuable tool for analyzing and quantifying the intertwined nature of coherence and entanglement.
The coherence concurrence is based on the \(\frac{d(d-1)}{2}\)-dimensional symmetric generalized GGM matrices, which serve as the symmetric generators of the \(\mathrm{SU}(d)\) Lie-algebra. These matrices are instrumental in capturing the multi-level structure of quantum states in higher-dimensional systems. Specifically, the symmetric GGM matrices are given by \(\Lambda_{s}^{i,j} = \ket{i}\bra{j} + \ket{j}\bra{i}\), where \(1 \leq i < j \leq d\). 

This measure satisfies all the stringent conditions required for a proper coherence measure: \(\boldsymbol{\rm C_1}\), \(\boldsymbol{\rm C_2}\), \(\boldsymbol{\rm C_2'}\) , and \(\boldsymbol{\rm C_3}\). The coherence concurrence is defined as the convex-roof of the \(l_1\)-norm of coherence, which is expressed mathematically as:
\begin{equation}
C_c(\rho) = \min_{\{p_i, |\psi_i\rangle\}} \sum_i p_i C_{l_1}\left(\ket{\psi_i}\bra{\psi_i}\right),
\end{equation}
where the minimization is performed over all possible pure state decompositions of the mixed state \(\rho\). Here, \(\rho\) is represented as \(\rho = \sum_i p_i \ket{\psi_i}\bra{\psi_i}\), where \(p_i\) denotes the probability of obtaining the \(i\)-th outcome, with \(p_i \geq 0\) and \(\sum_i p_i = 1\). In the case of any dimensional pure state, such as the mode-entangled neutrino systems considered in this study, it has been shown that the coherence concurrence \(C_c(\rho)\) simplifies to the \(l_1\)-norm of coherence. 

\subsection{Relative entropy of coherence}
Among the various distance measures of coherence, the relative entropy of coherence—also known as \textit{distillable coherence}—stands out as a fundamental canonical monotone. This measure holds particular significance as it can be operationally interpreted as the optimal rate at which one can distill a maximally coherent state from a given state \(\rho\) through incoherent CPTP operations \(\Xi_{\rm ICPTP}(\rho)\) in the asymptotic limit of many copies of the state \cite{Winter:2016bkw}.

The quantum relative entropy, which quantifies the metric distance between two density matrices \(\rho\) and \(\sigma\), is defined as:
\begin{equation}
    S(\rho\|\sigma) = {\rm Tr}\big[\rho(\log{\rho} - \log{\sigma})\big] = -{\rm Tr}\big[\rho(\log{\sigma})\big] - S(\rho),
\label{rel}
\end{equation}
where \(\sigma \in \mathcal{I}\) represents an incoherent quantum state, and \(S(\rho) \equiv -{\rm Tr}(\rho \log{\rho})\) denotes the von Neumann entropy of \(\rho(t)\).
By applying Eq. \eqref{rel}, we obtain the difference in relative entropies as:
\begin{equation}
    S(\rho_D\|\sigma) - S(\rho\|\sigma) = S(\rho) - S(\rho_D),
\end{equation}
where \(\rho_D(t)\) is the dephased state, constructed by retaining only the diagonal elements of the density matrix \(\rho(t)\).
The \textit{relative entropy of coherence} can thus be expressed in a closed-form as \cite{Baumgratz:2014yfv,Ding:2021zfi},
\begin{eqnarray}
    C_{\rm RE}(\rho) &=& S(\rho_{D}(t))-S(\rho(t)) \nonumber \\ &=& {\rm Tr}\big[\rho(t)\log{\rho(t)}-\rho_D(t)\log{\rho_D(t)}\big]\,.
    \label{relen}
\end{eqnarray}

In comparison to other measures, such as the $l_1$-norm, which captures coherence by summing the magnitudes of off-diagonal elements in the density matrix, the relative entropy of coherence offers a more profound and holistic understanding by encompassing the full informational content of the quantum state. Although the $l_1$-norm is advantageous for its simplicity and straightforward computation, it provides a more focused perspective, emphasizing only the contributions from off-diagonal components. On the other hand, the relative entropy of coherence assesses the overall distribution of coherence within the state and quantifies its potential as a quantum resource in practical applications. It gauges how distinguishable a quantum state is from its dephased, incoherent version, offering valuable insights into the state's effectiveness for quantum information processing tasks.

For our relevant pure states like $\ket{\nu_{\alpha}}\bra{\nu_{\alpha}}$ this measure is given as\footnote{Hereon, $\log$ represents binary logarithm, with base 2},
\begin{equation}
    C_{\rm RE}\left(\ket{\nu_{\alpha}}\bra{\nu_{\alpha}}\right)=S\left(\rho_{D\alpha}(t)\right)=-{\rm Tr}\big[\rho_{D\alpha}(t)\log{\rho_{D\alpha}(t)}\big]\,,
    \label{CofRE}
\end{equation}
where $\rho_{D\alpha}$ is the dephased state made out of the diagonal part of the pure density operator $\ket{\nu_{\alpha}}\bra{\nu_{\alpha}}$.
Using the density matrix in Eq. \eqref{den2}, we can write the relative entropy of coherence for the two-flavor oscillations of two-qubit neutrinos as,

\begin{equation}
\begin{aligned}
C_{\rm RE}^{2f}\left(\rho_{\alpha,2}^{4\times 4}\right) &= -\text{Tr}\big[ \rho_{D\alpha,2}^{4\times 4}(\theta,L) \log\rho_{D\alpha,2}^{4\times 4}(\theta,L)\big] \\
&= -\Big(|\widetilde{U}_{\alpha\alpha}(\theta,L)|^2 \log\big(|\widetilde{U}_{\alpha\alpha}(\theta,L)|^2\big) 
 + |\widetilde{U}_{\alpha\beta}(\theta,L)|^2 \log\big(|\widetilde{U}_{\alpha\beta}(\theta,L)|^2\big)\Big),
\label{relen2-1}
\end{aligned}
\end{equation}

where the survival probability $|\Tilde{U}_{\alpha\alpha}(\theta,L)|^2$ and the transition probability $|\Tilde{U}_{\alpha\beta}(\theta,L)|^2$ can be substituted using the coefficients from Eq. \eqref{2coef}. We then get the $C_{\rm RE}^{2f}$ in terms of  parameters of oscillation as,
\begin{widetext}
    \begin{eqnarray}
    C_{\rm RE}^{2f}\left(\rho_{\alpha,2}^{4\times 4}\right)&=&-\Big(\cos^4{\theta}+\sin^4{\theta}\Big)\log{\left(\cos^4{\theta}+\sin^4{\theta}+\frac{\sin^2{2\theta}}{2}\cos{\frac{\Delta m_{ji}^{2}}{2E}L}\right)}
    \nonumber\\ &-&
     \frac{\sin^2{2\theta}}{2}\cos{\frac{\Delta m_{ji}^{2}}{2E}L} \log\Bigg\{\cos^4{\theta}+\sin^4{\theta}+\frac{\sin^2{2\theta}}{2}\left(2\cos{\frac{\Delta m_{ji}^{2}}{2E}L}-1\right)\Bigg\}\nonumber\\ &-&
    \sin^2{2\theta}\log{\left(\sin{2\theta}\sin{\frac{\Delta m_{ji}^{2}}{4E}L}\right)}
   \,\nonumber
   \\&=& 
-P_{\alpha\alpha}^{2f}\log P_{\alpha\alpha}^{2f}
-P_{\alpha\beta}^{2f}\log P_{\alpha\beta}^{2f},
\label{relen2}
\end{eqnarray}

\end{widetext}

For three-flavor mode oscillations, when the initial neutrino flavor states $|\nu_{\alpha}\rangle$ are expressed in the tripartite three-qubit basis, we consider the three $8\times 8$ density matrices $\rho_{\alpha,3}^{8\times 8}(\theta_{ij},L)$ from Eq. \eqref{den32} and express the relative entropy of coherence as,

\begin{equation}
\begin{aligned}
C_{\mathrm{RE}}^{3f}\left(\rho_{\alpha,3}^{8 \times 8}\right) &= -\mathrm{Tr}\big[ \rho_{D\alpha,3}^{8 \times 8}(\theta_{ij}, L) \log \rho_{D\alpha,3}^{8 \times 8}(\theta_{ij}, L) \big] \\
&= -2 \sum_{\beta = e}^{\tau} |a_{\alpha \beta}(\theta_{ij}, L)|^2 \log |a_{\alpha \beta}(\theta_{ij}, L)|  \label{relen32}
\end{aligned}
\end{equation}

where the coefficients \( a_{\alpha\beta}(\theta_{ij}, L) \) for \( \alpha,\beta = \{e,\mu,\tau\} \) can be extracted from Eq. \eqref{a} and substituted into Eq. \eqref{relen32} to express it in terms of the oscillation parameters.  For the initial flavor $\alpha=e$, $C_{\rm RE}^{3f}\left(\rho_{e,3}^{8\times 8}\right)$ can be expressed as,

\begin{equation}
\begin{aligned}
C_{\mathrm{RE}}^{3f}\left(\rho_{e,3}^{8 \times 8}\right) &= -|a_{ee}(\theta_{ij}, L)|^{2} \log |a_{ee}(\theta_{ij}, L)|^{2} 
 - |a_{e\mu}(\theta_{ij}, L)|^{2} \log |a_{e\mu}(\theta_{ij}, L)|^{2} 
 - |a_{e\tau}(\theta_{ij}, L)|^{2} \log |a_{e\tau}(\theta_{ij}, L)|^{2},
\label{erel}
\end{aligned}
\end{equation}
 and for the initial flavor $\alpha=\mu$, $C_{\rm RE}^{3f}\left(\rho_{\mu,3}^{8\times 8}\right)$ can be expressed as,
\begin{equation}
\begin{aligned}
C_{\mathrm{RE}}^{3f}\left(\rho_{\mu,3}^{8 \times 8}\right) &= -|a_{\mu\mu}(\theta_{ij}, L)|^{2} \log |a_{\mu\mu}(\theta_{ij}, L)|^{2} 
 - |a_{e\mu}(\theta_{ij}, L)|^{2} \log |a_{e\mu}(\theta_{ij}, L)|^{2} 
 - |a_{\mu\tau}(\theta_{ij}, L)|^{2} \log |a_{\mu\tau}(\theta_{ij}, L)|^{2}.
\label{murel}
\end{aligned}
\end{equation}

The above expressions can be written in terms of probabilities as 
\begin{equation}
    C_{\mathrm{RE}}^{3f}\left(\rho_{\alpha,3}^{8 \times 8}\right) = -\sum_{\beta = e}^{\tau} P_{\alpha \beta}(\theta_{ij}, L) \log P_{\alpha \beta}(\theta_{ij}, L),
\end{equation}

\noindent \textbf{Coherence of formation }: The coherence of formation (or the \textit{Intrinsic randomness measure}), which equals coherence cost, is the minimal asymptotic rate of consuming maximally coherent pure state for preparing $\rho$ by incoherent operation \cite{Winter:2016bkw}. Operationally, it represents the minimum amount of coherence required to generate a given quantum state using incoherent operations, making it a fundamental concept in the resource theory of coherence. For a pure state $\ket{\psi}\bra{\psi}$ (where $|\psi\rangle=\sum_{i=1}^{d}a_i |i\rangle$) this measure is given as,
\begin{equation}
    R_{I}\left(\ket{\psi}\bra{\psi}\right)= S\left(\rho_D\right)=C_{\rm RE}\left(\ket{\psi}\bra{\psi}\right)\,,
    \label{CofRE2}
\end{equation}
where $\rho_D$ is the dephased state made out of the diagonal part of the pure state $\ket{\psi}\bra{\psi}$, and $C_{\rm RE}$ is the relative entropy of coherence. Being the convex-roof of the relative entropy of coherence, Eq. \eqref{CofRE} shows the equivalence of this measure with the relative entropy of coherence for pure states\footnote{This coherence measure is extended to mixed state $\rho$ by the convex roof construction (which inherently makes the measure satisfy the convexity property $\boldsymbol{\rm C_3}$) as $R_{I}\left(\rho\right)=\min_{\left\{p_i,|\psi_i\rangle\right\}}\sum_i p_i R_I\left(|\psi_i\rangle\right)$,
where the minimization is taken over all pure state decompositions and $\rho=\sum_i p_i \ket{\psi_i}\bra{\psi_i}$.}. This measure can be physically interpreted as states with stronger coherence would, therefore, indicate larger randomness in measurement outcomes, and vice versa.
Both the relative entropy of coherence and the coherence of formation are known to satisfy two additional conditions on top of the four conditions mentioned at the start of this section. One of these conditions is the condition of \textit{uniqueness} for pure states is shown in Eqs. \eqref{CofRE} and \eqref{CofRE2}. The other condition is \textit{additivity} under tensor products, i.e., $C\left(\rho\otimes\sigma\right)=C\left(\rho\right)+C\left(\sigma\right)$, which are also satisfied by the logarithmic coherence rank.

\subsection{Schatten-$p$-norm-based functionals}
The Schatten-$p$-norm $\|\cdot\|_p$  of any linear-bounded operator $A$ is given by\footnote{Here the case we are interested in is the $p=1$ trace class norm, other important values of $p$ here include $p=2$ and $p\rightarrow\infty$ corresponding to the Hilbert-Schmidt norm and the operator norm respectively.},
\begin{equation}
     \|A\|_p=\left({{\rm Tr}}\left[ \left(A^{\dagger}A\right)^{\frac{p}{2}} \right] \right)^{\frac{1}{p}}; \quad\, p\in[1,\infty)\,,
\end{equation}
The two classes of functionals used for measuring coherence, based on the aforementioned Schatten norm are given for $p\geq 1$ as,
\begin{eqnarray}
     C_{p}=\min_{\delta\epsilon\mathcal{I}} \|\rho-\delta\|_p ,\quad\, \Tilde{C}_{p}= \|\rho-\rho_{D}\|_p\,,
    \label{sch}
\end{eqnarray}
As per \cite{Cui:2020},
\begin{itemize}
    \item  $C_{p=1}$ is not a valid coherence measure under incoherent operations, strictly incoherent operations and genuinely incoherent operations\footnote{Genuinely incoherent operations are a subset of strictly incoherent operations, and the latter are a subset of incoherent operations.}. 
    \item  $\Tilde{C}_{p=1}$ is a valid coherence measure under both strictly incoherent operations and genuinely incoherent operations but NOT under incoherent operations.
    \item  Neither $C_{p>1}$ nor $\Tilde{C}_{p>1}$ is a valid coherence measure under any of the three sets of incoherent operations.
\end{itemize}

From the above points, it is reasonable to pursue the calculation of the second class of functionals based on the Schatten-$p=1$-norm given by,
\begin{equation}
   \Tilde{C}_{p=1}\left(\rho\right)={{\rm Tr}}\left[ \left(\left(\rho-\rho_{D}\right)^{\dagger}\left(\rho-\rho_{D}\right)\right)^{\frac{1}{2}} \right]\,.
\end{equation}
The Schatten-$p=1$-norm-based functional offers a way to assess the coherence properties of neutrino states by measuring the extent to which the actual state differs from its dephased counterpart. This difference indicates the level of coherence, which is essential for understanding the entanglement and evolution of neutrino oscillations. Unlike other measures, such as the \(l_1\)-norm of coherence, $ \Tilde{C}_{p=1}$ captures the overall distance between the density matrix and its dephased version, offering a more comprehensive perspective on coherence. While the \(l_1\)-norm primarily accounts for the sum of off-diagonal elements, the trace norm evaluates the entire matrix, making it a more encompassing measure in certain scenarios.

For the two-flavor mode entangled two-qubit neutrino system, we can take the density matrix in Eq. \eqref{den2} and write this functional as,

\begin{equation}
\begin{aligned}
\widetilde{C}_{p=1}^{2f}\left(\rho_{\alpha,2}^{4\times 4}\right) &= \mathrm{Tr}\Big[ \big\{\left(\rho_{\alpha,2}^{4\times 4}(\theta,L) - \rho_{D\alpha,2}^{4\times 4}(\theta,L)\right)^\dagger  \left(\rho_{\alpha,2}^{4\times 4}(\theta,L) - \rho_{D\alpha,2}^{4\times 4}(\theta,L)\right)\big\}^{1/2} \Big] \\
&= 2 |\widetilde{U}_{\alpha\alpha}(\theta,L)||\widetilde{U}_{\alpha\beta}^*(\theta,L)|,
\end{aligned}
\end{equation}
Now substituting the coefficients from Eq. \eqref{2coef} in the above equation, we finally get,
\begin{widetext}
\begin{equation}
\begin{aligned}
\Tilde{C}_{p=1}^{2f}\left(\rho_{\alpha,2}^{4\times 4}\right)
&=2 \sin{2\theta}\sin{\frac{\Delta m_{ji}^{2}}{4E}L}
\sqrt{\left(\cos^4{\theta}+\sin^4{\theta}
+\frac{\sin^2{2\theta}}{2}\cos{\frac{\Delta m_{ji}^{2}}{2E}L}\right)}
\\
&=2\sqrt{P_{\alpha\alpha}^{2f}P_{\alpha\beta}^{2f}}\,,
\end{aligned}
\end{equation}

which is identical to $C_{l_1}^{2f}\left(\rho_{\alpha,2}^{4\times 4}\right)$ for two-flavor case.

\end{widetext}

Now, for the three-flavor mode entangled three-qubit neutrino system, we can take the density matrices from Eq. \eqref{den32} and write this functional as,
\begin{widetext}

\begin{eqnarray}
&&\Tilde{C}_{p=1}^{3f}\left(\rho_{\alpha,3}^{8\times 8}\right)={{\rm Tr}}\left[ \left\{\left(\rho_{\alpha,3}^{8\times 8}(\theta_{ij},L)-\rho_{D\alpha,3}^{8\times 8}(\theta_{ij},L)\right)^{\dagger}\left(\rho_{\alpha,3}^{8\times 8}(\theta_{ij},L)-\rho_{D\alpha,3}^{8\times 8}(\theta_{ij},L)\right)\right\}^{\frac{1}{2}} \right]\nonumber\\&=& \big|a_{\alpha\tau}(\theta_{ij},L)\big|\sqrt{\big|a_{\alpha\mu}(\theta_{ij},L)\big|^2+\big|a_{\alpha e}(\theta_{ij},L)\big|^2}+\big|a_{\alpha\mu}(\theta_{ij},L)\big|\sqrt{\big|a_{\alpha\tau}(\theta_{ij},L)\big|^2+\big|a_{\alpha e}(\theta_{ij},L)\big|^2}\nonumber\\&+&\quad\big|a_{\alpha e}(\theta_{ij},L)\big|\sqrt{\big|a_{\alpha\mu}(\theta_{ij},L)\big|^2+\big|a_{\alpha\tau}(\theta_{ij},L)\big|^2}\nonumber\\&=&
\sqrt{P_{\alpha\tau}(\theta_{ij},L)\left(P_{\alpha\mu}(\theta_{ij},L)+P_{\alpha e}(\theta_{ij},L)\right)}+\sqrt{P_{\alpha\mu}(\theta_{ij},L)\left(P_{\alpha\tau}(\theta_{ij},L)+P_{\alpha e}(\theta_{ij},L)\right)}\nonumber\\&+&\sqrt{P_{\alpha e}(\theta_{ij},L)\left(P_{\alpha\mu}(\theta_{ij},L)+P_{\alpha\tau}(\theta_{ij},L)\right)}\,,
\label{eq:snorm}
\end{eqnarray}

\end{widetext}
where, in the final step, we have used the relation
\(
P_{\alpha\beta}(\theta_{ij},L)
=
|a_{\alpha\beta}(\theta_{ij},L)|^2
\)
to express \(\Tilde{C}_{p=1}^{3f}\left(\rho_{\alpha,3}^{8\times 8}\right)\) entirely in terms of the three-flavor oscillation probabilities given in Eq.~\eqref{moda2}. Thus, for any initial flavor \(\alpha=e,\mu,\tau\), the measure \(\Tilde{C}_{p=1}^{3f}\) can be directly evaluated from the corresponding oscillation probabilities \(P_{\alpha e}(\theta_{ij},L)\), \(P_{\alpha\mu}(\theta_{ij},L)\), and \(P_{\alpha\tau}(\theta_{ij},L)\).

\subsection{Geometric coherence as a fidelity-based measure}
The geometric coherence is defined in general between two density operators $\rho$ and $\delta$ as,
\begin{equation}
    C_{g}\left(\rho\right)=1-\max_{\delta\in\mathcal{I}}F\left(\rho,\delta\right)\,,
\end{equation}
where the fidelity is expressed as the square of the Uhlmann fidelity as $F\left(\rho,\delta\right)=\big[{\rm Tr}\sqrt{\rho^{1/2}\delta\rho^{1/2}}\big]^2$ is unity for $\rho=\delta$, and non-negative i.e., satisfying the condition $\boldsymbol{\rm C_1}$. As the fidelity is non-decreasing under CPTP maps, it makes $C_{g}\left(\rho\right)$ non-increasing under the action of $\Xi_{\rm ICPTP}$ and thus monotonic, satisfying $\boldsymbol{\rm C_2}$. It also satisfies $\boldsymbol{\rm C_2'}$ and $\boldsymbol{\rm C_3}$ and is thus a proper measure.

Geometric coherence quantifies the distinguishability of a quantum state from the nearest incoherent state in a geometric sense, where ``geometric" refers to the mathematical distance or proximity between states in the high-dimensional Hilbert space. This measure is based on fidelity, which evaluates the overlap between the quantum state $\rho$ and the closest incoherent state $\delta$, providing a nuanced perspective on how far a state is from being incoherent. The geometric coherence captures this distance, making it particularly useful in assessing the robustness of coherence under various quantum transformations. This approach is valuable in practical scenarios like quantum computing and quantum metrology, where understanding how coherence is maintained or lost under operations is crucial. Unlike other coherence measures that focus on the total amount of coherence, geometric coherence offers an intuitive and geometrically grounded way to assess a quantum state's coherence, particularly in terms of its resilience to noise and decoherence.

For pure states, the fidelity is given as $F\left(\rho,\delta\right)={\rm Tr}\left(\rho\delta\right)=\sum_{i=1}^{d}\rho_{ii}\delta_{i}\leq\max_i{\{\rho_{ii}\}}$, and thus the geometric coherence for pure states in $d$-dimensional Hilbert space becomes,
\begin{eqnarray}
C_{g}\left(\rho\right) &=& 1-\max_i{\{\rho_{ii}\}} \nonumber\\ &=& 1-\frac{1}{d}-\frac{d-1}{d}\sqrt{1-\frac{d}{d-1}\left(1-\sum_{i}\rho_{ii}^2\right)}\,.
    \label{gc}
\end{eqnarray}
\begin{widetext}
Now from the above Eq. \eqref{gc}, we can calculate the geometric coherence for two-flavor neutrino oscillations when neutrinos are mapped to two-qubit states ($d=4$) using the $4\times 4$ density matrix $\rho_{\alpha,2}^{4\times 4}\left(\theta,L\right)$ from Eq. \eqref{den2} as,
\begin{eqnarray}
    C_{g}^{2f}\left(\rho_{\alpha,2}^{4\times 4}\right)&=&1-\max_i{\{(\rho_{\alpha,2}^{4\times 4})_{ii}\}}\nonumber\\ &=&1-\max{\{|\Tilde{U}_{\alpha\alpha}(\theta,L)|^2,|\Tilde{U}_{\alpha\beta}(\theta,L)|^2\}}\nonumber\\
    &=&\frac{3}{4}-\frac{3}{4}\sqrt{1-\frac{4}{3}\left(1-\left(|\Tilde{U}_{\alpha\alpha}(\theta,L)|^4 + |\Tilde{U}_{\alpha\beta}(\theta,L)|^4\right)\right)}\,,
    \label{gc2-1}
\end{eqnarray}
Putting the coefficients from Eq. \eqref{2coef} in the above equation Eq. \eqref{gc2-1}, we get,


\begin{eqnarray}
C_{g}^{2f}\left(\rho_{\alpha,2}^{4\times 4}\right)
&=&
1-\max{\Bigg\{
\cos^4{\theta}+\sin^4{\theta}
+\frac{\sin^2{2\theta}}{2}
\cos{\frac{\Delta m_{ij}^{2}L}{2E}},
\,
\sin^2{2\theta}
\sin^2{\frac{\Delta m_{ij}^{2}L}{4E}}
\Bigg\}}
\nonumber\\
&=&
1-\max\left\{
P_{\alpha\alpha}^{2f},
P_{\alpha\beta}^{2f}
\right\}
\nonumber\\
&=&
\frac{1}{2}
\left[
1-
\left|
P_{\alpha\alpha}^{2f}
-
P_{\alpha\beta}^{2f}
\right|
\right].
\label{gc2}
\end{eqnarray}
\end{widetext}

When the three-flavor neutrino states are mapped to the three-qubit system given by the density matrix $\rho_{\alpha,3}^{8\times 8}(\theta_{ij},L)$ in Eq. \eqref{den32}, the geometric coherence takes the form,
\begin{equation}
\begin{aligned}
C_{g}^{3f}\left(\rho_{\alpha,3}^{8 \times 8}\right) &= 1 - \max_{i}\big\{(\rho_{\alpha,3}^{8 \times 8})_{ii}\big\} \\
&= 1 - \max\big\{|a_{\alpha e}(\theta_{ij}, L)|^2, |a_{\alpha\mu}(\theta_{ij}, L)|^2,  |a_{\alpha\tau}(\theta_{ij}, L)|^2\big\} \\
&= 1 - \max\big\{P_{\alpha e}(\theta_{ij}, L), P_{\alpha\mu}(\theta_{ij}, L), P_{\alpha\tau}(\theta_{ij}, L)\big\},
\label{gc31}
\end{aligned}
\end{equation}
where, we use the relation
\(
P_{\alpha\beta}(\theta_{ij},L)
=
|a_{\alpha\beta}(\theta_{ij},L)|^2
\)
to express the geometric coherence entirely in terms of the three-flavor oscillation probabilities given in Eq.~\eqref{moda2}. Thus, for any initial flavor \(\alpha=e,\mu,\tau\), the geometric coherence is determined by the largest among the probabilities \(P_{\alpha e}(\theta_{ij},L)\), \(P_{\alpha\mu}(\theta_{ij},L)\), and \(P_{\alpha\tau}(\theta_{ij},L)\).

\noindent \textbf{Fidelity of coherence}: Another related measure called the fidelity of coherence is a measure based on the fidelity-induced distance between two states $\rho$ and $\delta$ and can be given as,
\begin{equation}
    C_{f}\left(\rho\right)=1-\sqrt{\max_{\delta\in\mathcal{I}}F\left(\rho,\delta\right)}\,,
\end{equation} 
where again the non-decreasing property of fidelity $F\left(\rho,\delta\right)$ under CPTP maps, makes the fidelity-induced distance $1-\sqrt{F\left(\rho,\delta\right)}$ and in turn $C_{f}\left(\rho\right)$ monotonic. This measure satisfies the condition $\boldsymbol{\rm C_2}$, but it has been shown in \cite{Shao:2014} using amplitude damping-like qubit ICPTP map, $C_{f}\left(\rho\right)$ is a functional that does not qualify as a proper measure of coherence since it cannot be classified as ``strong'' monotone, i.e., it violates the strong monotonicity condition 
$\boldsymbol{\rm C_2'}$\footnote{In general, $C_{f}\left(\rho\right)\neq1-\sqrt{{\rm Tr}\sqrt{\rho^{1/2}\rho_D\rho^{1/2}}}$ since the dephased state $\rho_D$ is non necessarily optimal for the fidelity of coherence, making the subselection process hard to verify.}.

Although both geometric coherence and fidelity of coherence rely on fidelity, they approach the relationship between quantum states differently. Geometric coherence evaluates the highest fidelity across all incoherent states, whereas fidelity of coherence incorporates a square root function, potentially altering its sensitivity to specific quantum transformations.

\subsection{Logarithmic coherence rank}
The logarithmic coherence rank $C_\mathcal{L}\left(|\psi\rangle\right)$
 serves as a critical measure in the resource theory of coherence, offering unique insights into the structural complexity of quantum states.
For pure states, the coherence number is the same as the coherence rank, and for mixed states, the coherence number is the smallest possible maximal coherence rank in any decomposition of the mixed state. For any pure state $|\psi\rangle$ the logarithmic coherence rank is defined as \cite{Killoran},
\begin{equation} C_\mathcal{L}\left(|\psi\rangle\right)=\log_2{R_C\left(|\psi\rangle\right)}=\log_2\{\min \sum_{|j\rangle\epsilon\hat{\mathcal{O}}}\lambda_j |j\rangle\}; \quad\, \hat{\mathcal{O}}\subseteq\mathcal{O}\,,
    \label{loga}
\end{equation}
where $\lambda_j$ are non-zero complex coefficients, and $R_C\left(|\psi\rangle\right)$ is the coherence rank that represents the minimal number of the incoherent states in the fixed orthogonal basis ${\mathcal{O}}$ in the decomposition of $|\psi\rangle$\footnote{For pure states in any $d$-dimensional Hilbert space, the coherence rank satisfies the inequality $2\leq R_C\leq d$}. This measure characterizes the logarithm of the minimal number of incoherent states in the fixed orthogonal basis $\mathcal{O}$ (in Hilbert space $\mathcal{H}$) in such a decomposition of the pure state $|\psi\rangle$, thus revealing the multilevel nature of coherence. 
For a coherent pure state living in a $d$-dimensional Hilbert space, the minimum coherence rank is $R_{C,min}\left(|\psi\rangle\right)=2$, while for the maximally coherent state, $R_{C,max}\left(|\psi\rangle\right)=d$. Hence, the logarithmic coherence rank satisfies the inequality $1\leq C_\mathcal{L}\left(|\psi\rangle\right)\leq \log_2{d}$.

As the coherence rank \(R_C\left(|\psi\rangle\right)\) provides a direct count of the minimal number of non-zero coefficients (incoherent states) required in the decomposition of a pure state, the logarithmic coherence rank then offers a more nuanced view by expressing this count in logarithmic terms, thereby reflecting the multilevel nature of coherence in terms of information content. This logarithmic scaling is particularly useful in comparing coherence across systems of different dimensions. 

In comparison to other coherence measures like the \(l_1\)-norm or the relative entropy of coherence, the logarithmic coherence rank uniquely emphasizes the structural intricacies of the quantum state by counting the number of non-zero coefficients in its decomposition. While the \(l_1\)-norm and relative entropy are primarily concerned with quantifying the total coherence present in the state, the logarithmic coherence rank provides deeper insights into how coherence is distributed among the various components of the quantum state.

In practical terms, the logarithmic coherence rank’s emphasis on the minimal number of significant components in a quantum state’s decomposition can provide deeper insights into the operational capabilities of quantum systems, particularly in tasks like quantum computation, where the distribution of coherence directly impacts algorithmic efficiency. Further, in the context of quantum computing, a higher logarithmic coherence rank suggests an increased ability to encode and manipulate information, as a larger number of non-zero coefficients generally implies greater availability of quantum resources. Moreover, its logarithmic nature aligns well with the scaling of resources in quantum communication protocols, offering a valuable perspective in the study of high-dimensional quantum systems.

The definition of the measure in Eq. \eqref{loga} can be extended to mixed states by the standard convex-roof construction, and we can thus define the logarithmic coherence number as \cite{Xi:2019},
\begin{equation}
    C_\mathcal{L}\left(\rho\right)=\min_{\left\{p_i,|\psi_i\rangle\right\}}\sum_{i}p_i C_\mathcal{L}\left(|\psi\rangle\right)\,,
\end{equation}
where the minimization is taken over all pure state decompositions of $\rho=\sum_i p_i |\psi_i\rangle\langle\psi_i|$, wherein $p_i$ is the probability of obtaining the $i$-th outcome. In \cite{Xi:2019}, $C_\mathcal{L}$ was established as a proper measure of coherence.

To begin with, in our two-qubit mode-entangled neutrino system, the decomposition of the two-flavor evolution, as given in Eq. \eqref{2evol}, reveals that two out of the four possible coefficients are non-zero in the neutrino flavor state's decomposition in the Bell-basis. Notably, the coefficients corresponding to the Bell states \( |0\rangle \otimes |0\rangle \) and \( |1\rangle \otimes |1\rangle \) are zero, as evident from Eq. \eqref{2evol}. Consequently, the coherence rank for this system is \( R_C\left(\rho_{\alpha,2}^{4\times 4}\right) = 2 \), and the logarithmic coherence rank is \( C_\mathcal{L}^{2f}\left(\rho_{\alpha,2}^{4\times 4}\right) = \log_2{2} = 1 \). Similarly, for the three-flavor evolution of the mode-entangled three-qubit neutrino system, there are three non-zero coefficients out of the possible eight in the decomposition within the tripartite basis of the three-qubit system (given in Eq. \eqref{3evol2}). 

This analysis highlights that the (logarithmic) coherence rank consistently equals the (logarithm of) the minimal number of non-zero coefficients in the multipartite basis of any dimensional entangled system. For a state with \( k \) non-zero coefficients \( \lambda_j \) in Eq. \eqref{loga}, the logarithmic coherence rank is \( C_\mathcal{L}\left(|\psi\rangle\right) = \log_2{k} \). For maximally coherent states \( |\psi_M\rangle \), this becomes \( C_\mathcal{L}\left(|\psi_M\rangle\right) = \log_2{d} \), where \( d \) is the dimension of the Hilbert space. Therefore, in the two-qubit system, the maximum logarithmic coherence rank is    
\( C_{\mathcal{L},max}\left(\rho_{\alpha,2}^{4\times 4}\right) = \log_2{4} = 2 \);  and in the three-qubit system, it is \( C_{\mathcal{L},max}\left(\rho_{\alpha,3}^{8\times 8}\right) = \log_2{8} = 3 \). 

\section{Behavior of Coherence Measures}

In this section, we demonstrate the dynamical behaviour of various quantum coherence measures for neutrino oscillations within the three-flavor mode entanglement framework. The values of the oscillation parameters used in our analysis are listed in Table~\ref{tab:osc_params}.

\begin{table}[H]
\centering
\caption{Oscillation parameters used in the numerical analysis for normal ordering~\cite{Esteban:2024eli}.}
\label{tab:osc_params}
\begin{tabular}{c c}
\hline
Parameter & Value  \\
\hline
$\theta_{12}$ & $33.82^\circ$ \\
$\theta_{13}$ & $8.61^\circ$ \\
$\theta_{23}$ & $48.3^\circ$ \\
$\Delta m_{21}^{2}$ & $7.42\times 10^{-5}~{\rm eV}^{2}$ \\
$\Delta m_{31}^{2}$ & $2.51\times 10^{-3}~{\rm eV}^{2}$ \\
\hline
\end{tabular}
\end{table}

\begin{figure}[H]
    \centering
\includegraphics[width=0.46\textwidth]{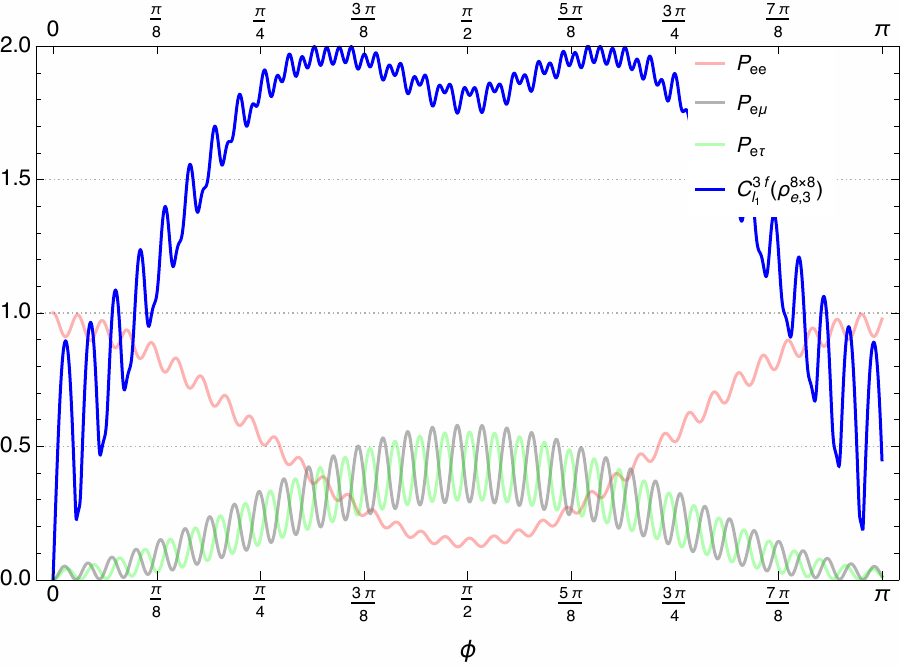} 
\includegraphics[width=0.46\textwidth]{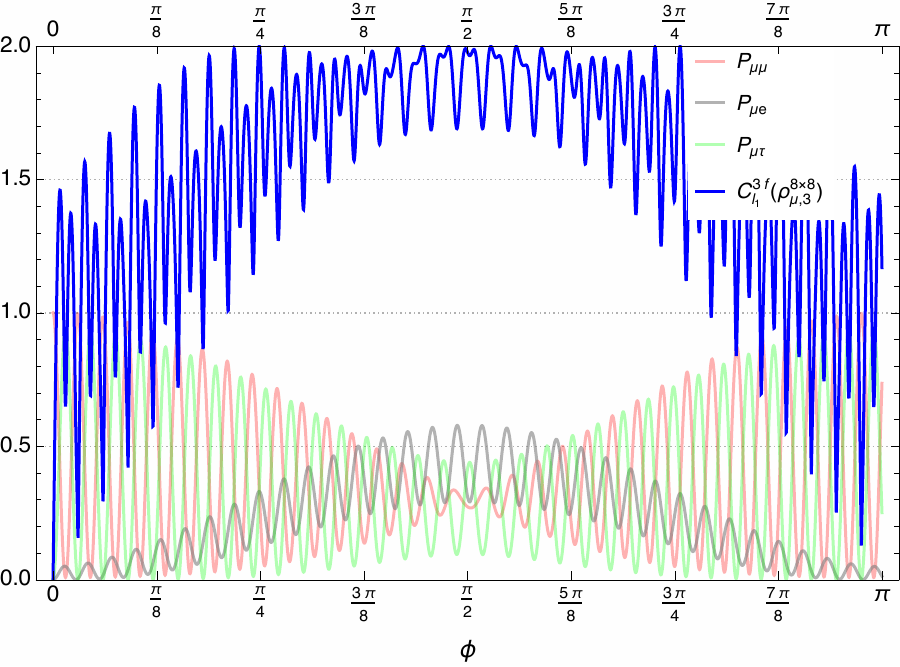}
  \caption{The figure shows the \(l_1\)-norm of coherence \(C_{l_1}^{3f}\), together with the corresponding survival and transition probabilities.  All quantities are plotted as functions of \(\phi\). The left panel corresponds to the initial electron neutrino state and the right panel corresponds to initial muon neutrino. The analysis is performed within the framework of three-flavor neutrino oscillations, where the neutrino state is treated as a three-mode system and mapped onto the tripartite basis of a three-qubit system.
}
\label{fig:l1}
\end{figure}

The $l_1$-norm of coherence for the initial electron- and muon-neutrino states in the three-qubit mode-entangled representation can be written in terms of oscillation probabilities from Eq.~\eqref{eq:l1prob}. The $l_1$-norm of coherence in the three-qubit framework depends on the propagation distance $L$ and the neutrino energy $E$. The evolution of $C_{l_1}^{3f}\left(\rho_{\alpha,3}^{8\times 8}\right)$ for $\alpha=e,\mu$ is shown in Fig.~\ref{fig:l1}, together with the corresponding survival and transition probabilities. The left panel corresponds to the initial electron-neutrino state, while the right panel corresponds to the initial muon-neutrino state. To capture both solar- and atmospheric-scale behavior without choosing a specific baseline or neutrino energy, we plot the probabilities and all coherence measures in this section as functions of the dimensionless phase
\begin{equation}
\phi=\frac{\Delta m_{21}^{2}L}{4E}.
\end{equation}

For both initial flavor neutrinos, the $l_1$-norm of coherence increases when the initial flavor survival probability decreases and the state becomes distributed among the other flavor modes. When the initial survival probability is large, the coherence is comparatively smaller. As the transition probabilities increase, the coherence remains higher because the state becomes more shared among the flavor modes. This can be directly seen from Eq.~\eqref{eq:l1prob}, where the measure depends on the products of oscillation probabilities for different flavors. It can also be seen from both panels of Fig.~\ref{fig:l1} that the $l_1$-norm approaches its maximum when the corresponding probabilities become almost equal i.e.

\[
P_{\alpha e}\simeq P_{\alpha\mu}\simeq P_{\alpha\tau}\simeq \frac{1}{3},
\qquad \alpha=e,\mu .
\]

Here, the arguments of the probabilities are dropped for compactness. 

\begin{figure*} [htb]
    \centering
    \includegraphics[width=0.46\textwidth]{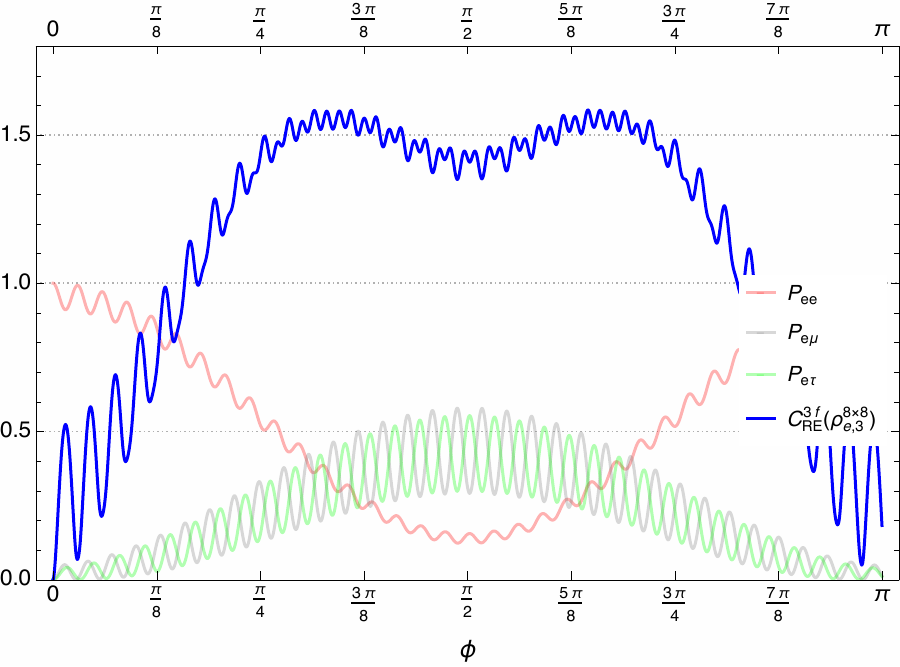}   \includegraphics[width=0.46\textwidth]{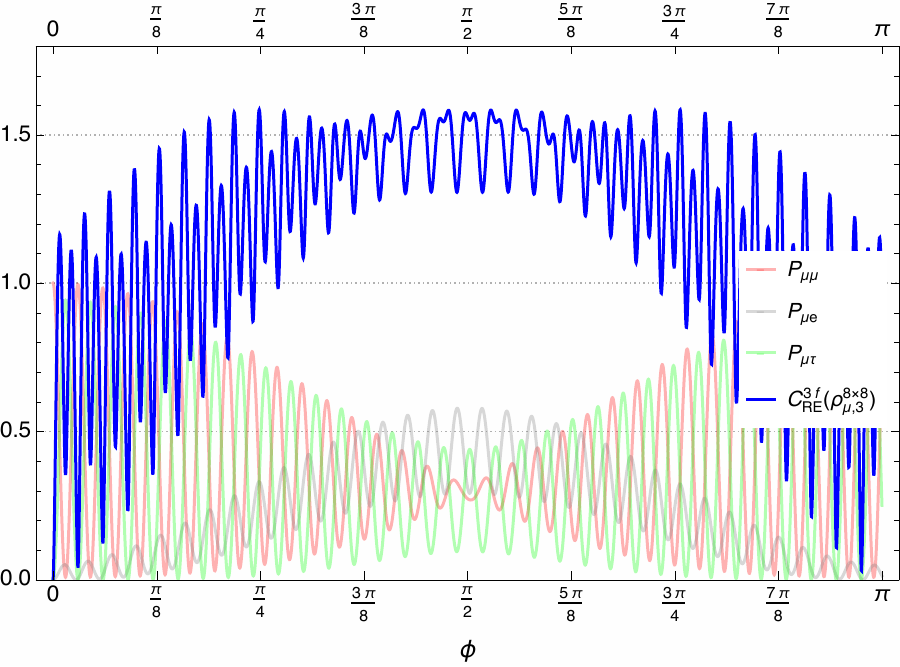}

\caption{The relative entropy of coherence, \(C_{\rm RE}^{3f}\), together with the associated survival and transition probabilities, is plotted as a function of \(\phi\) for three-flavor neutrino oscillations. The left (right) panel corresponds to an initially produced electron (muon) neutrino, described in the tripartite three-qubit representation.}
    \label{fig:re}
\end{figure*}

The relationship between the relative entropy of coherence and the survival and transition probabilities for both initial muon-neutrino and electron-neutrino states is illustrated in Fig. \ref{fig:re}. The values of \( C_{\rm RE}^{3f}\left(\rho_{\alpha,3}^{8\times 8}\right) \) correspond to the scenario where these neutrino states are mapped onto the tripartite basis of a three-qubit system and are given in Eqs.~\eqref{erel} and~\eqref{murel}. A high relative entropy of coherence indicates that the quantum state possesses a substantial amount of coherence, signifying that the state is far from being incoherent and maintains a significant degree of quantum superposition. This implies that the state can maintain and utilize quantum superpositions effectively.

Relative entropy suggests a similar behaviour to that of $l_1$-norm. This again suggests that the coherence is enhanced when the neutrino is less likely to remain in its initial flavor state, making \( C_{\rm RE}^{3f}\left(\rho_{\alpha,3}^{8\times 8}\right) \) a useful indicator of flavor redistribution during oscillations. As the neutrino state becomes more spread among different flavor modes, the coherence measure increases, reflecting the stronger quantum superposition present in the system. For the initial \( \nu_e \) state, the peak value of \( C_{\rm RE}^{3f}\left(\rho_{e,3}^{8\times 8}\right) \), which is approximately 1.5. For the initial \( \nu_\mu \) state, the maximum of \( C_{\rm RE}^{3f}\left(\rho_{\mu,3}^{8\times 8}\right) \) is also approximately same as of electron neutrino initial state.

Furthermore, the plots reveal that the coherence, as measured by \( C_{\rm RE}^{3f}\left(\rho_{\alpha,3}^{8\times 8}\right) \) and \( C_{l_1}^{3f}\left(\rho_{\alpha,3}^{8\times 8}\right) \), remains significant over a large part of the oscillation cycle. Even when the survival probability is relatively high, the coherence does not necessarily vanish, indicating that the system can retain a finite level of flavor-mode superposition during propagation. The multiple peaks observed in the relative entropy of coherence show that the neutrino system experiences varying degrees of coherence as it evolves.

\begin{figure*}[htb]
    \centering
\includegraphics[width=0.46\textwidth]{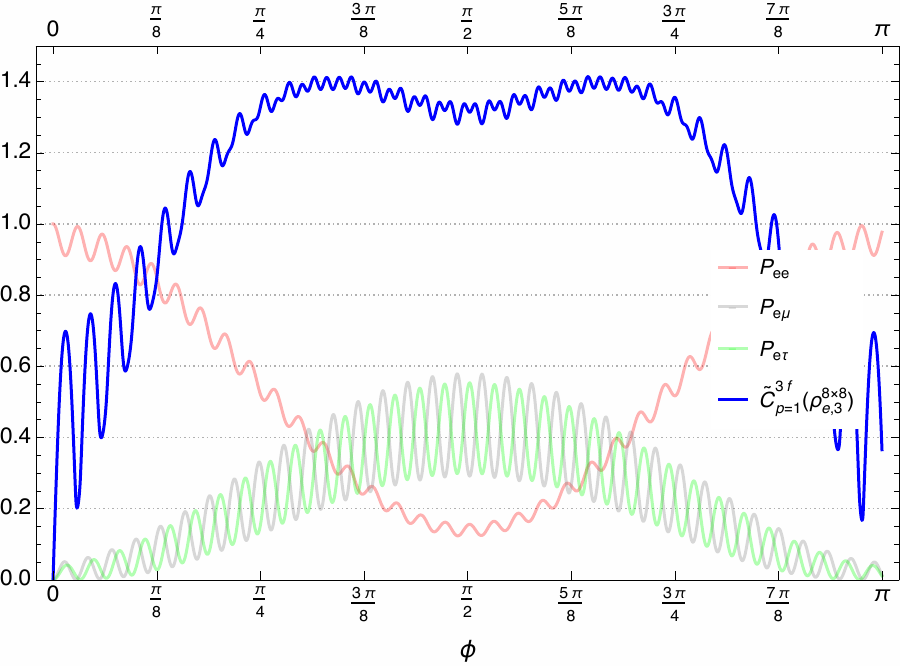}   \includegraphics[width=0.46\textwidth]{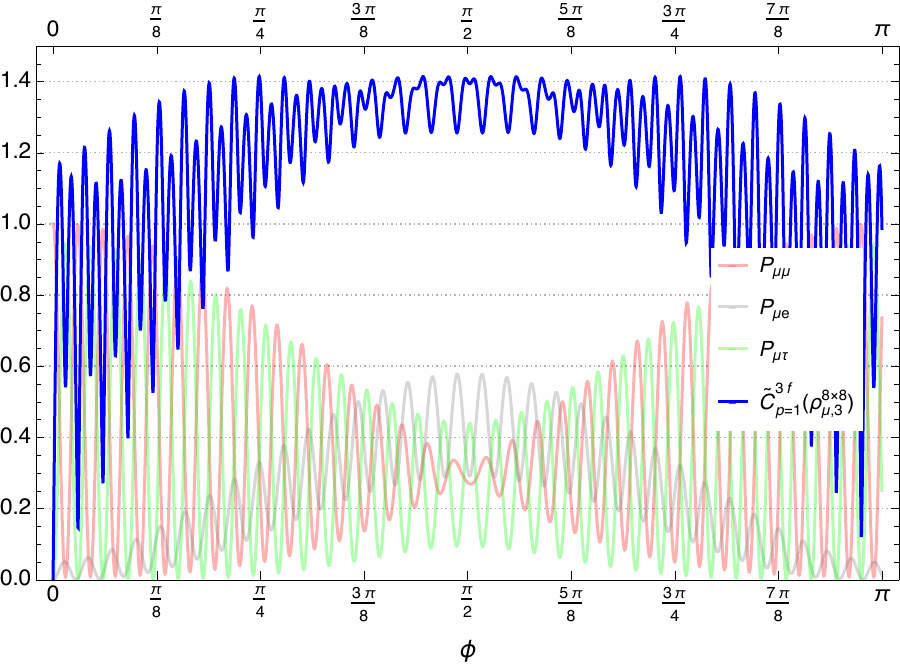}
    \caption{The Schatten-\(p=1\)-norm-based functional, \(\Tilde{C}_{p=1}^{3f}\), together with the corresponding flavor-transition probabilities, is shown as a function of \(\phi\) for three-flavor neutrino oscillations. The left (right) panel corresponds to an initially produced \(\nu_e\) (\(\nu_\mu\)), where the flavor state is represented in the tripartite three-qubit basis.}
    \label{fig:sh}
\end{figure*}

Fig. \ref{fig:sh} corresponds to the dynamics of the Schatten-\( p = 1 \)-norm-based functional \( \Tilde{C}_{p=1}^{3f}\left(\rho_{\alpha,3}^{8\times 8}\right) \) for three-flavor neutrino oscillations, mapped to a three-qubit system using Eq.~\eqref{eq:snorm} 
. It can be seen from the plot that the Schatten-\( p = 1 \)-norm-based functional \( \Tilde{C}_{p=1}^{3f}\left(\rho_{\alpha,3}^{8\times 8}\right) \) exhibits oscillatory behavior, similar to other coherence measures, reflecting the quantum coherence dynamics of the neutrino system. For both initial electron and muon neutrino states, the coherence measure oscillates between a minimum value near zero and a maximum value slightly above 1.4.

\begin{figure}[htb]
\includegraphics[width=0.49\textwidth]{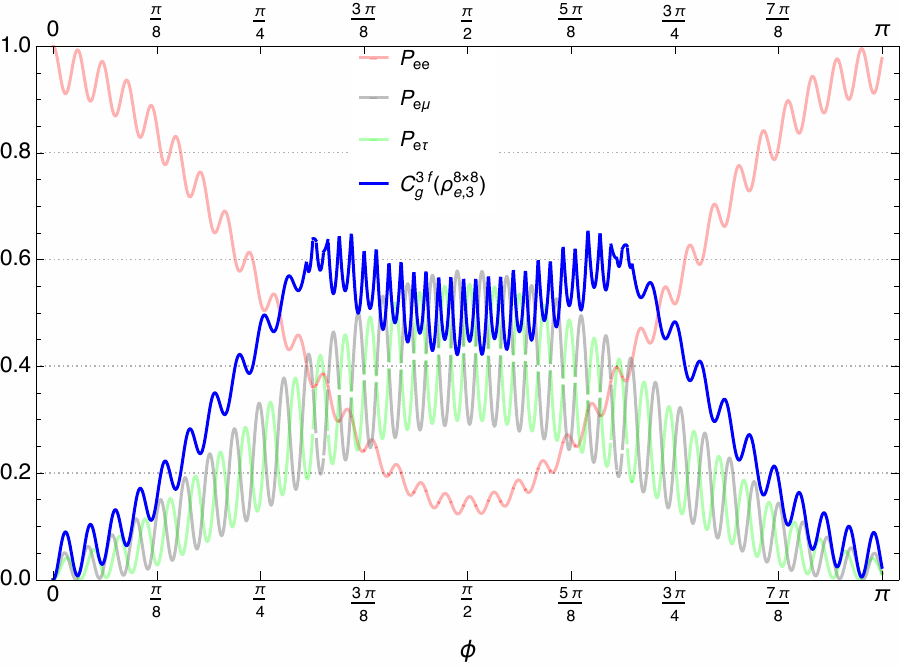} 
\includegraphics[width=0.49\textwidth]{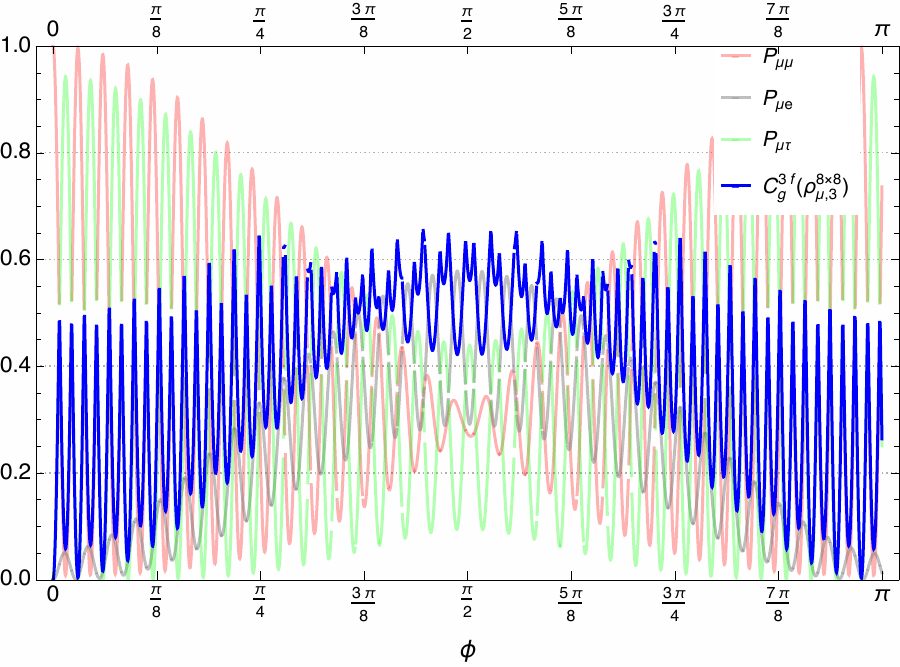}
    \caption{Displayed are the geometric coherence $C_{g}^{3f}$ and the corresponding survival and transition probabilities as functions of $\phi$ for three-flavor neutrino oscillations. The left (right) panel corresponds to an initial electron (muon) neutrino, with the neutrino state represented as a three-mode system mapped onto a tripartite three-qubit basis.}
    \label{fig:gc}
\end{figure} 
As with the \( l_1 \)-norm and the relative entropy of coherence, the higher values of \( \Tilde{C}_{p=1}^{3f}\left(\rho_{\alpha,3}^{3\times 3}\right) \) tend to correspond to the intervals where the survival probability \( P_{\alpha\alpha} \) is low and it reaches to the peak when all corresponding probabilities are nearly equal. The pattern of coherence dynamics is similar for both initial electron and muon neutrino states, though the specific locations of the peaks and troughs differ slightly due to the different mixing parameters involved in each case.

The plots for geometric coherence \( C_{g}^{3f}\left(\rho_{\mu,3}^{8\times 8}\right) \) and \( C_{g}^{3f}\left(\rho_{e,3}^{8\times 8}\right) \), shown in Fig.~\ref{fig:gc}, illustrate the coherence dynamics of neutrino oscillations for both initial states, respectively, mapped onto the tripartite basis of the three-qubit system. The geometric coherence also shows a behavior similar to the $l_1$-norm and the relative entropy of coherence. This measure of coherence is defined in Eq.~\eqref{gc31} in terms of neutrino oscillation probabilities. It is evident that it also becomes large when the state is equally shared among the flavors. It reaches its peak, \(C_g^{3f}=2/3\simeq 0.6\), when all probabilities are close to $1/3$. This shows that the system reaches a maximally shared flavor state.

For all four coherence measures discussed in this section, the differences in the rapid modulations observed for the initial electron- and muon-neutrino states arise from the distinct flavor-transition structures encoded in the PMNS matrix. In particular, the \(\nu_\mu\) state is more strongly influenced by the atmospheric mixing sector through the large value of \(\theta_{23}\), leading to more pronounced oscillatory modulations, while the overall magnitude of the coherence measures remains comparable for both initial states.

\section{Conclusions}

In this comprehensive study, we have explored the profound interplay between quantum coherence and neutrino oscillations, framing neutrino flavor states within multi-mode quantum systems and analyzing them through the rigorous lens of quantum information theory. Recognizing quantum coherence as the cornerstone of all quantum correlations, our investigation has delved into its manifestation and evolution in neutrino systems. By providing a detailed methodology for calculating all relevant coherence measures in both two-flavor and three-flavor mixing scenarios, we have equipped the field with a methodological framework to facilitate future experimental and theoretical endeavors aimed at harnessing neutrinos for quantum technologies. Our analysis extended beyond traditional measures of quantum coherence, such as the \(l_1\)-norm and relative entropy of coherence, to include a suite of sophisticated metrics like the Schatten-\(p\)-norm-based functionals, geometric coherence and logarithmic coherence rank.

Each measure provided unique insights, taken together, the different coherence measures provide complementary information about the flavor-mode coherence generated during neutrino oscillations. The \(l_1\)-norm of coherence serves as a baseline measure for quantifying the degree of superposition within neutrino states and is closely linked to the oscillation probabilities.  
The trace norm distance measure reduces to the \(l_1\)-norm of coherence for the two-flavor mode-entangled oscillations of neutrino states with Bell-diagonal structure. Moreover, the robustness of coherence and coherence concurrence also simplify to the \(l_1\)-norm of coherence for the neutrino states under consideration. 

The relative entropy of coherence provides a more refined characterization by measuring the distance of the system from an incoherent state, while the coherence of formation, or intrinsic randomness measure, reduces to the relative entropy of coherence for the pure neutrino states considered here. The Schatten-\(p\)-norm-based functional is sensitive to the structure of the entangled system and also becomes large when the probability distribution among the flavor modes becomes more balanced when the neutrino state is mapped onto the tripartite basis of the three-qubit system. The geometric coherence provides a geometric perspective on the quantumness of neutrino oscillations and reaches its maximum  when all probabilities are almost equal, showing that the system approaches a maximally shared flavor state. 

The Logarithmic Coherence Rank captures the structural complexity of the quantum state in terms of the number of non-zero coefficients appearing in its decomposition in a chosen reference basis. In particular, it increases logarithmically with the number of coherently populated flavor modes participating in the oscillation dynamics. Consequently, the logarithmic coherence rank takes a lower value for the two-qubit neutrino system and a higher value for the three-qubit flavor-mode system, since the minimum number of non-zero coefficients in the decomposition of the flavor state is two in the former case and three in the latter. Hence, this measure provides a useful characterization of the multipartite structure of flavor-mode coherence in neutrino oscillations.

Therefore, all these measures show that coherence in neutrino oscillations is directly connected with the redistribution of the initial flavor state among different flavor modes. We also show that all measures show richer structure in three-flavor than the two-flavor oscillation framework as these can be expressed in terms of neutrino oscillation probabilities, which are directly relevant to neutrino oscillation experiments. This behavior is evident for both initial electron-neutrino and muon-neutrino states. This work advances our understanding of the quantum mechanical nature of neutrino oscillations. We note that extending the present analysis to a wave-packet framework would allow for a spatial characterization of neutrino oscillations and provide a more realistic description of the dynamical evolution of quantum coherence during propagation.

\section*{Acknowledgements}
This work is dedicated to our friend and colleague Prof. Ashutosh K. Alok, who passed away during the final stages of manuscript preparation. We thank the referees for their careful reading of the manuscript and for their valuable comments, which have helped us improve the clarity and presentation of the paper. NRSC would like to express sincere thanks to the Galileo Galilei Institute for Theoretical Physics, Florence, for their generous hospitality while completing part of this work.
NRSC also wishes to express deep gratitude to the Department of Physics ``E.R. Caianiello" and INFN, Salerno, for their generous hospitality and support during the early stages of this project.

\end{document}